\documentclass{aa}

\newcommand{\vnh}{\hat {\mbox{\bf n}}}

\newcommand{\vx}{\mbox{\bf {x}}}
\newcommand{\vy}{\mbox{\bf {y}}}
\newcommand{\vr}{\mbox{\bf {r}}}
\newcommand{\vk}{\mbox{\bf {k}}}

\newcommand{\lnu}{L_{\nu}}

\newcommand{\dkthree}{\frac{d\vk}{(2\pi)^3}}

\newcommand{\ex}{\times10^}

\newcommand{\s}{$\star$}

\usepackage{lscape}
\usepackage{graphicx}
\usepackage{natbib}
\usepackage{longtable}
\usepackage{supertabular}
\bibpunct{(}{)}{;}{a}{}{,}
\usepackage{txfonts}
\begin{document}

   \title{Carbon monoxide line emission as a CMB foreground: tomography of the star-forming universe with different spectral resolutions}
   \titlerunning{Carbon monoxide line emission as a CMB foreground}
   \authorrunning{M. Righi et al.}

   \author{M. Righi
          \inst{1},
          C. Hern\'andez-Monteagudo
          \inst{1},
          \and
          R.A. Sunyaev\inst{1,2}
          }

   \offprints{C. Hern\'andez-Monteagudo,\\ e-mail: chm@mpa-garching.mpg.de}

   \institute{Max-Planck-Institut f\"ur Astrophysik, 
              Karl-Schwarzschild-Str. 1, 85748 Garching, Germany
\and
              Space Research Institute (IKI), 
              Profsoyuznaya 84/32, Moscow, 117810, Russia
             }

   \date{Received ...; ...}

 
  \abstract
   {The rotational lines of carbon monoxide and the fine structure lines of CII and of the most abundant metals, emitted during the epoch of enhanced star formation in the universe, are redshifted in the frequency channels where the present-day and future CMB experiments are 
sensitive.}
   {We estimate the contribution to the CMB angular power spectrum by the emission in such lines in merging star-forming galaxies.}
   {We used the Lacey-Cole approach to characterize the distribution of the merging halos, together with a parametrization for the star formation rate in each of them. Using observational data from a sample of local, low-redshift, and high-redshift objects, we calibrated the luminosity in each line as a function of the star formation rate.}
   {We show that the correlation term arising from CO line emission is a significant source of foreground for CMB in a broad range of frequencies (in particular in the $20-60$~GHz band) and for $1000<l<8000$, corresponding to angular scales smaller than 10 arcminutes. Moreover, we demonstrate that observing with different spectral resolutions will give the possibility of increasing the amplitude of the signal up to two 
orders of magnitude in $C_l$ and will help separate the line contribution from practically \emph{all} other foreground sources and from the primary fluctuations themselves, since these show no significant dependence on the spectral resolution.}
   {We propose to perform observations with varying spectral bandwidths ($10^{-3}<\Delta\nu/\nu_{\rm obs}<10^{-1}$) as a new tool to construct a tomography of the universe, by probing different redshift slices with varying thickness. This should yield new constraints on the regions responsible for the metal enrichment in the universe and on their clustering pattern and will lead to new hints about the reionization epoch and the cosmological parameters, including $\sigma_8$.}


   \maketitle


\section{Introduction}\label{sec:intro}
The nucleosynthesis in the first generation of stars is able to produce significant amounts of carbon and oxygen  \citep{hw02,yoshida06}. This bring our attention to the carbon monoxide rotational lines originating from the star-forming halos in the early universe. Recent observations show that these lines are very bright in the most distant quasars and radiogalaxies at redshift $z\sim4$ \citep[see e.g.][]{greve05} and also in the star-forming galaxies in our vicinity \citep{gao01,weiss05a,bayet06,baan08}. They are also the brightest radio lines in our Galaxy \citep{wright91,fixsen99}.

In this paper we are making an attempt to compute the angular power spectrum of the foreground fluctuations due to emission in such lines from merging star-forming galaxies. The main difficulty of this task is related to the presence of the continuum emission of dust at the same frequencies. To separate the line contribution, we propose observing in several spectral bands with resolutions in the range $\Delta\nu/\nu_{\mathrm{obs}}=10^{-1}-10^{-3}$. We demonstrate that with this technique it will be possible to increase the amplitude of the signal due to the lines
emission significantly (over one order of magnitude), while the contribution from other foregrounds generated by continuum emission (dust, radio sources, SZ effects in clusters of galaxies, and the primordial fluctuations themselves) does not depend on the spectral resolution of the observations. Observations of the power spectrum of CO line emission with high spectral resolution allow narrow slices of the universe to be probed at different redshifts, with a thickness $\Delta z/z\sim\Delta\nu/\nu_{\mathrm{obs}}$. Performing our computations for different spectral resolutions and proving the strength of this method, we recognized that an analogous approach is used in demonstrating the results of numerical simulations of the large-scale structure of the universe. In that case the most effective contrast is achieved using slices with a very small thickness ($\sim15$ Mpc) compared to the huge dimensions of the box ($\sim$ Gpc\footnote{\tt http://www.mpa-garching.mpg.de/galform/millennium/}). The observations in the CO lines can use comparable $\Delta\nu/\nu_{\mathrm{obs}}$ to increase the amplitude of the signal and to separate it from other continuum sources.

It is rather difficult, at present, to obtain a reliable theoretical estimate of the luminosity of different CO lines in the star-forming halos \citep{silk97,combes99,greve08}, and for this reason we use existing observational data to calibrate our model. We consider local merging galaxies and luminous infrared galaxies, together with a sample of high-redshift, sub-millimeter galaxies, observed at $z\sim2-4$. Using these samples, we calibrate the luminosity of the different lines on the star formation rate of the object, assuming a linear scaling. We follow the same approach as outlined in our previous paper \citep{righi08}, where we presented a model for computing the fluctuations due to dust emission in merging star-forming halos. This model, based on the extended Press-Schechter formalism \citep[EPS,][]{lc93}, allowed us to obtain a statistical description of the distribution of the halos as a function of their star formation rate. With these two independent ingredients, we can compute the power spectrum of angular fluctuations arising from the emission in the CO lines, at any frequency and for any spectral
resolution. This approach will allow the integral properties of the population of weak star-forming merging galaxies to be measured in a broad redshift range: by calibrating with observational data obtained from
different galaxy samples, we can describe the history of formation of giant molecular clouds and model the subsequent process of CO enrichment of the IGM throughout cosmic history.

The beauty and the strength of the proposed method of observing the angular fluctuations in the narrow line emission is connected to the fact that, for a given line at a fixed observing frequency, we detect the radiation from a slice of the universe with given $\Delta z$. The data on the cosmic star formation history \citep[the so-called \emph{Madau plot},][]{madau96,madau98,hopkins06} shows that the bulk of the star formation activity in the universe takes place in a very broad redshift range. Here we are considering the star-forming objects in the range~$0\la z\la10$, corresponding to $\nu\simeq10-115$~GHz for the first CO transition. This is close to the WMAP and PLANCK LFI's detectors \citep{WMAP,LFI}. Future ground-based experiments operating in the centimeter band and able to perform observations in many relatively narrow spectral channels ($10^{-3}<\Delta\nu/\nu_{\rm obs}<10^{-1}$) might open the possibility of separating the contribution of line emission from any other continuum source using the method proposed in this paper. More accurate semi-analytical models \citep[see e.g.][]{granato04,lacey08,cole08,reed08} might allow the predictions to be extended to even higher redshifts and to correspondingly lower frequencies.\\

Higher CO transitions are still bright up to $\sim350$~GHz and are therefore interesting for the PLANCK HFI's bands \citep{HFI}, as well as for ACT and SPT \citep{ACT,SPT}. The same spectral band can cover different slices of the universe, and from this point of view we will see that the $10-40$~GHz region is the most interesting one for the CO emission, because only the first one or two transitions (and the corresponding redshift slices) contribute significantly to the $C_l$'s in this band. We present, however, the frequency (and redshift) distribution for all the CO transitions up to $J=7$, which will permit recognition of the redshift range that gives the main contribution to the signal.

According to the same model, we also compute the differential source counts curve for the emission in every line. The cumulated signal from many weak sources might contribute more to the power spectrum due to the clustering. Therefore, in addition to the simple Poisson fluctuations, we computed the correlation signal taking into account the regions of higher merging activity, corresponding to the positions where the future clusters and super-clusters of galaxies will appear.

From the same sample of objects that we used to calibrate our model, we see that the emission from the CN, HCN, HNC, and HCO$^+$ molecules, with resonant frequency close to the CO ones (within $\sim20\%$), do not add more than $10\%$ to the amplitude of the CO signal \citep{baan08}. At higher frequencies, the dominant contribution comes from the CII 158~$\mu$m line, while other atomic lines are much weaker. Here, however, we are interested not in the absolute luminosity of the lines, but in their ratio to the primordial fluctuation and to dust continuum emission for the same angular scales, in the same spectral band and with a given $\Delta\nu/\nu_{\mathrm{obs}}$. This argument makes the first two CO transitions more important than the (more luminous) higher transitions, given their contribution to the power spectrum of angular fluctuations relative to the primordial CMB signal and to the foreground from extragalactic sources emitting in the sub-millimeter band due to the presence of large amounts of dust. There are no galactic foregrounds that can mimic a similar dependence on the spectral resolution to the signal from CO. The first CO transitions are usually saturated, and their ratio follows the same Rayleigh-Jeans law of the CMB at low frequencies. At the same time, the dust particles have very low opacity at low frequencies making their effect weaker with decreasing frequency, compared with CO~(1-0) and (2-1) transitions.

We use the EPS formalism because it is relatively simple and permits us to demonstrate the key results of the proposed method. The deviations that the EPS introduces with respect to the predictions of numerical simulation \citep{cole08} are still smaller than the uncertainties in the observational sample that we use for the calibration of the CO line luminosities. Such uncertainties are mainly due to the data for the star formation rates: in the low-redshift sample we use only the IRAS data, combined with the Kennicutt relation, to estimate the star formation rate. The galaxies in the high-redshift sample, on the other hand, often harbor an AGN and this is a source of further uncertainty. An additional difficulty comes from the different angular resolution of the CO and infrared observations.

Future observations with different bandwidths will measure the dependence of the CO contribution on spectral resolution. More precise numerical models can refine the predictions of our simple model. In our view, it is important to propose a new method of directly measuring the rate of enrichment of the universe with CO not in the very rare and extremely bright objects (as QSO and radiogalaxies at $z\sim4-6$), but in the most abundant and less bright objects, which cannot be detected individually.

\begin{figure*}
\centering
  \includegraphics[width=\textwidth]{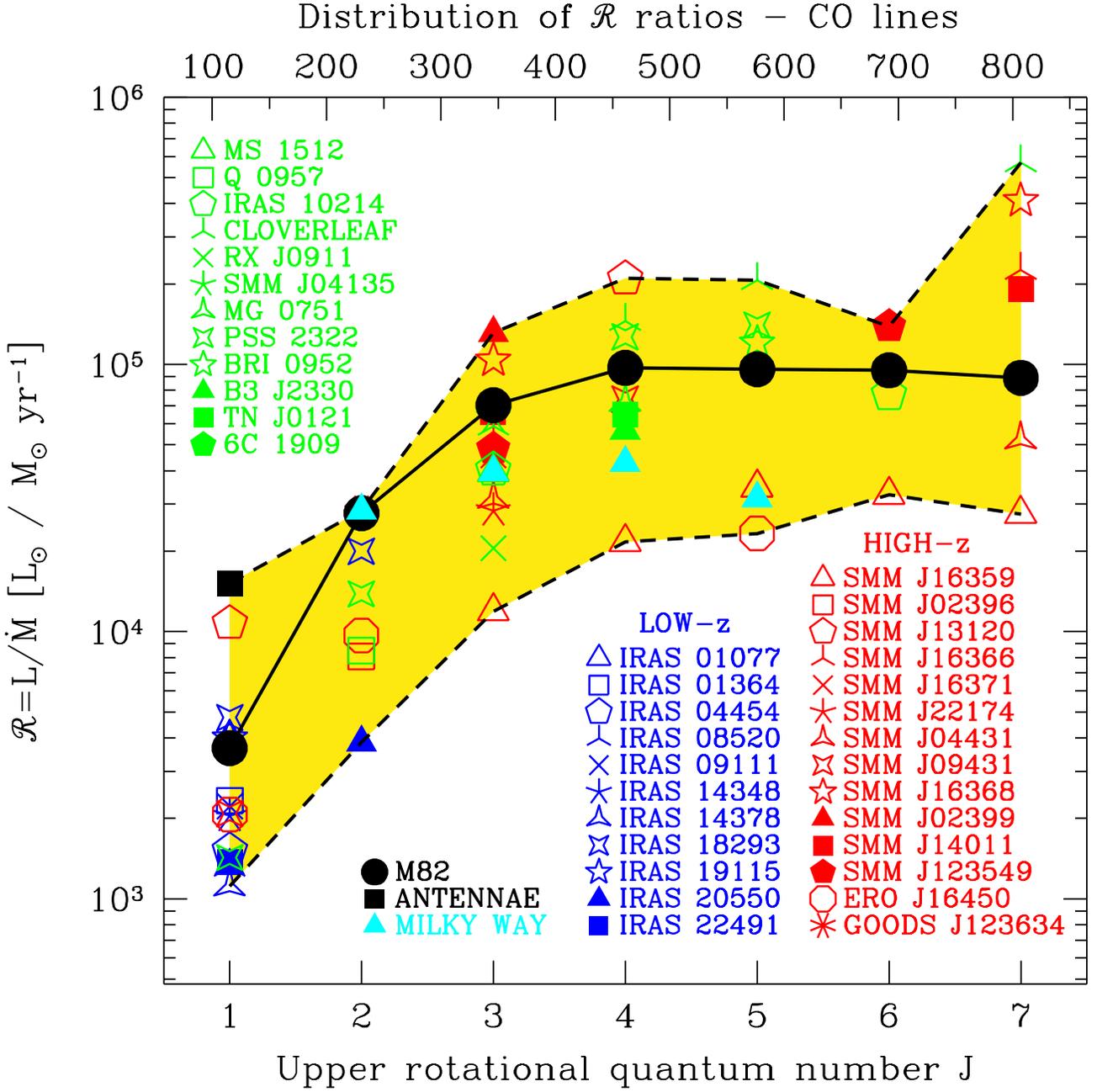}
  \caption{The distribution of the $\mathcal{R}=L/\dot{M}$ ratios as a function of the upper rotational quantum number of the CO transitions (the corresponding frequency is shown in the upper axis in GHz). Three different samples are considered: low-redshift IRAS galaxies (blue points), high-redshift sub-millimeter galaxies (red points), and high-redshift radiogalaxies and QSOs \citep[][green points]{greve05,solomonvdb05}. M82 and the Antennae are shown in black. Our Galaxy \citep{wright91,cox00} is represented with the cyan triangles. See Tables~\ref{tab:COlowz}~and~\ref{tab:COhighz}  for details. The black solid line is SED for M82, for which we have data for the full set of lines. Dashed lines represent the upper and lower limits for $\mathcal{R}$ in the sample.}
  \label{fig:COplot}
\end{figure*}

\begin{table*}
\caption{Luminosities and $\mathcal{R}$ ratios (in boldface) of the CO lines for the local and low-redshift sample of galaxies.}
\scriptsize
\label{tab:COlowz}
\begin{minipage}{\textwidth}
\begin{center}
\begin{tabular}{lccccccccccc}
\hline 
\hline 
Object name			& Type	&$d_{\rm L}$& SFR			& CO (1-0) 		& CO (2-1) 		& CO (3-2) 		& CO (4-3) 		& CO (5-4) 		& CO (6-5) 		& CO (7-6) 		\\
				&		&		&			& $115.3$ GHz	& $230.5$ GHz	& $345.8$ GHz	& $461.0$ GHz	& $576.3$ GHz	& $691.5$ GHz	& $806.7$ GHz	\\
				& 		& [Mpc]	&[$M_{\odot}$/yr]	&[$L_{\odot}$]	&[$L_{\odot}$]	&[$L_{\odot}$]	&[$L_{\odot}$]	&[$L_{\odot}$]	&[$L_{\odot}$]	&[$L_{\odot}$]	\\
\hline 
M82				& SB		& 3.63 	& 10.1		& $3.7\ex{4}$ a	& $2.8\ex{5}$ a	& $7.1\ex{5}$ a	& $9.8\ex{5}$ a	& $9.7\ex{5}$ a	& $9.6\ex{5}$ a	& $9.0\ex{5}$ a	\\
				&		&		&			& $\bf3.7\ex{3}$	& $\bf2.8\ex{4}$	& $\bf7.0\ex{4}$	& $\bf9.7\ex{4}$	& $\bf9.6\ex{4}$	& $\bf9.5\ex{4}$	& $\bf8.9\ex{4}$	\\
\multicolumn{4}{r}{M82 $L_{\rm{dust}}/L_{\rm CO}$ ratio:}		& 			& $0.005$		& $0.009$ 		& $0.02$		& $0.05$		& $0.12$		& $0.24$		\\
\hline 
Antennae			& SB		& 21.8	& 11.8		& $1.8\ex{5}$ b	&			&			&			&			&			&			\\
				&		&		&			& $\bf1.5\ex{4}$	&			&			&			&			&			&			\\
Milky Way			& 		& -		& 3.0 		&			& $8.5\ex{4}$ c	& $1.2\ex{5}$ c	& $1.3\ex{5}$ c	& $9.5\ex{4}$ c	&			&			\\
				&		&		&			&			& $\bf2.8\ex{4}$	& $\bf4.0\ex{4}$	& $\bf4.3\ex{4}$	& $\bf3.2\ex{4}$	&			&			\\
IRAS 01077-1707		&		& 145 	& 72.9		& $2.9\ex{5}$ d	&			&			&			&			&			&			\\
				&		&		&			& $\bf4.0\ex{3}$	&			&			&			&			&			&			\\
IRAS 01364-1042		&		& 207 	& 98.4		& $2.3\ex{5}$ d	&			&			&			&			&			&			\\
				&		&		&			& $\bf2.3\ex{3}$	&			&			&			&			&			&			\\
IRAS 04454-4838		&		& 220 	& 105			& $1.6\ex{5}$ d	&			&			&			&			&			&			\\
				&		&		&			& $\bf1.5\ex{3}$	&			&			&			&			&			&			\\
IRAS 08520-6850		&		& 205 	& 98.4 		& $1.3\ex{5}$ d	&			&			&			&			&			&			\\
				&		&		&			& $\bf1.3\ex{3}$	&			&			&			&			&			&			\\
IRAS 09111-1007		&		& 243 	& 171			& $3.8\ex{5}$ d	&			&			&			&			&			&			\\
				&		&		&			& $\bf2.2\ex{3}$	&			&			&			&			&			&			\\
IRAS 14348-1447		& SB		& 388 	& 341			& $6.9\ex{5}$ d	&			&			&			&			&			&			\\
				&		&		&			& $\bf2.0\ex{3}$	&			&			&			&			&			&			\\
IRAS 14378-3651		&		& 315 	& 242			& $2.7\ex{5}$ d	&			&			&			&			&			&			\\
				&		&		&			& $\bf1.1\ex{3}$	&			&			&			&			&			&			\\
IRAS 18293-3413		&		& 80.6 	& 110			& $5.3\ex{5}$ d	& $2.2\ex{6}$ d	&			&			&			&			&			\\
				&		&		&			& $\bf4.8\ex{3}$	& $\bf2.0\ex{4}$	&			&			&			&			&			\\
IRAS 19115-2124		&		& 216 	& 127			& $5.1\ex{5}$ d	&			&			&			&			&			&			\\
				&		&		&			& $\bf4.0\ex{3}$	&			&			&			&			&			&			\\
IRAS 20550+1656		&		& 136 	& 127			& $1.7\ex{5}$ d	& $4.8\ex{5}$ d	&			&			&			&			&			\\
				&		&		&			& $\bf1.3\ex{3}$	& $\bf3.8\ex{3}$	&			&			&			&			&			\\
IRAS 22491-1808		& SB		& 350 	& 220			& $3.0\ex{5}$ d	&			&			&			&			&			&			\\
				&		&		&			& $\bf1.4\ex{3}$	&			&			&			&			&			&			\\
\hline 
\end{tabular}
\end{center}
References: (a) \citet {weiss05a}, (b) \citet{gao01}, (c) \citet{wright91}, (d) \citet{baan08}.
\end{minipage}
\end{table*}

\begin{table*}
\caption{Luminosities and $\mathcal{R}$ ratios (in boldface) of the CO lines for the high-redshift sample of galaxies.}
\scriptsize
\label{tab:COhighz}
\begin{minipage}{\textwidth}
\begin{center}
\begin{tabular}{lccccccccccc}
\hline 
\hline 
Object name		& Type	& $z$		& SFR			& CO (1-0) 		& CO (2-1) 		& CO (3-2) 		& CO (4-3) 		& CO (5-4) 		& CO (6-5) 		& CO (7-6) 		\\
			&		&		&			& $115.3$ GHz	& $230.5$ GHz	& $345.8$ GHz	& $461.0$ GHz	& $576.3$ GHz	& $691.5$ GHz	& $806.7$ GHz	\\
			& 		& 		&[$M_{\odot}$/yr]	&[$L_{\odot}$]	&[$L_{\odot}$]	&[$L_{\odot}$]	&[$L_{\odot}$]	&[$L_{\odot}$]	&[$L_{\odot}$]	&[$L_{\odot}$]	\\
\hline 
SMM J16359\s	& SB		& 2.52	& 500 a		&			&			& $6.0\ex{6}$ b-d& $1.1\ex{7}$ b	& $1.7\ex{7}$ b	& $1.6\ex{7}$ b	& $1.4\ex{7}$ b	\\
			&		&		&			&			&			& $\bf1.2\ex{4}$	& $\bf2.2\ex{4}$	& $\bf3.4\ex{4}$	& $\bf3.2\ex{4}$	& $\bf2.8\ex{4}$	\\
SMM J02396\s	& AGN		& 1.06	& 975 e		&			& $7.9\ex{6}$ f	&			&			&			&			&			\\
			&		&		&			&			& $\bf8.1\ex{3}$	&			&			&			&			&			\\
SMM J13120		& AGN		& 3.41	& 810 g		& $1.0\ex{7}$ g	& 			&			& $1.7\ex{8}$ f	&			&			&			\\
			&		&		&			& $\bf1.2\ex{4}$	&			&			& $\bf2.1\ex{5}$	&			&			&			\\
SMM J16366		& SB		& 2.45	& 1455 i		& 			&			& $7.7\ex{7}$ f,h	&			&			&			& $3.3\ex{8}$ h	\\
			&		&		&			&			&			& $\bf5.3\ex{4}$	&			&			&			& $\bf2.3\ex{5}$	\\
SMM J16371		& SB+AGN	& 2.38	& 877 i		& 			&			& $4.0\ex{7}$ f	&			&			&			&			\\
			&		&		&			&			&			& $\bf4.6\ex{4}$	&			&			&			&			\\
SMM J22174		& SB		& 3.10	& 1800 e		&			&			& $5.1\ex{7}$ f	&			&			&			&			\\
			&		&		&			&			&			& $\bf2.8\ex{4}$	&			&			&			&			\\
SMM J04431\s	& SB+AGN	& 2.51	& 450 e		&			&			& $1.4\ex{7}$ h,j	&			&			&			& $2.4\ex{7}$ h	\\
			&		&		&			&			&			& $\bf3.1\ex{4}$	&			&			&			& $\bf5.3\ex{4}$	\\
SMM J09431\s	& SB+AGN	& 3.35	& 1200 g		& $<2.5\ex{6}$ g	&			&			& $8.9\ex{7}$ h,j	&			&			&			\\
			&		&		&			& $\bf<2.1\ex{3}$	&			&			& $\bf7.4\ex{4}$	&			&			&			\\
SMM J16368		& SB+AGN	& 2.38	& 897 i		&			&			& $9.3\ex{7}$ h,j	&			&			&			& $3.7\ex{8}$ h,j	\\
			&		&		&			&			&			& $\bf1.0\ex{5}$	&			&			&			& $\bf4.1\ex{5}$	\\
SMM J02399\s	& SB+AGN	& 2.80	& 500 k		&			&			& $6.6\ex{7}$ k,l	&			&			&			&			\\
			&		&		&			&			&			& $\bf1.3\ex{5}$	&			&			&			&			\\
SMM J14011\s	& SB		& 2.56	& 360 e		&			&			& $2.4\ex{7}$ l,m	&			&			&			& $6.9\ex{7}$ m	\\
			&		&		&			&			&			& $\bf6.7\ex{4}$	&			&			&			& $\bf1.9\ex{5}$	\\
SMM J123549		& SB+AGN	& 2.20	& 1163 n		&			&			& $5.6\ex{7}$ h	&			&			& $1.6\ex{8}$ h	&			\\
			&		&		&			&			&			& $\bf4.8\ex{4}$	&			&			& $\bf1.4\ex{5}$	&			\\
ERO J16450		& SB		& 1.44	& 1539 o		& $3.2\ex{6}$ o	& $1.5\ex{7}$ d	&			&			& $3.6\ex{7}$ p	&			&			\\
			&		&		&			& $\bf2.1\ex{3}$	& $\bf9.7\ex{3}$	& 			&			& $\bf2.3\ex{4}$	&			&			\\
GOODS J123634	& SB		& 1.22	& 950	q		&			& $2.6\ex{7}$ q	&			&			&			&			&			\\
			&		&		&			&			& $\bf2.8\ex{4}$	&			&			&			&			&			\\
MS 1512\s		& LBG		& 2.73	& 15 e		&			&			& $5.9\ex{5}$ r	&			&			&			&			\\
			&		&		&			&			&			& $\bf4.0\ex{4}$	&			&			&			&			\\
Q 0957\s		& QSO		& 1.41	& 900	e		&			& $7.6\ex{6}$ s	&			&			&			&			&			\\
			&		&		&			&			& $\bf8.5\ex{3}$	&			&			&			&			&			\\
IRAS F10214\s	& QSO		& 2.29	& 540 e		&			&			& $2.2\ex{7}$ t-w &			&			& $4.2\ex{7}$ v	&			\\
			&		&		&			&			&			& $\bf4.0\ex{4}$	&			&			& $\bf7,7\ex{4}$	&			\\
CLOVERLEAF\s	& QSO		& 2.56	& 810 e		&			&			& $4.8\ex{7}$ x-aa& $1.2\ex{8}$ z	& $1.7\ex{8}$ z	&			& $4.6\ex{8}$ z	\\
			&		&		&			&			&			& $\bf5.9\ex{4}$	& $\bf1.5\ex{5}$	& $\bf2.1\ex{5}$	&			& $\bf5.7\ex{5}$	\\
RX J0911\s		& QSO		& 2.80	& 345 e		&			&			& $7.1\ex{6}$ ab	&			&			&			&			\\
			&		&		&			&			&			& $\bf2.1\ex{4}$	&			&			&			&			\\
SMM J04135\s	& QSO		& 2.84	& 3600 e		&			&			& $2.3\ex{8}$ ab	&			&			&			&			\\
			&		&		&			&			&			& $\bf6.4\ex{4}$	&			&			&			&			\\
MG 0751\s		& QSO		& 3.20	& 435	e		&			&			& 			& $3.2\ex{7}$ ac	&			&			&			\\
			&		&		&			&			&			&			& $\bf7.2\ex{4}$	&			&			&			\\
PSS J2322\s		& QSO		& 4.11	& 1800 e		& $2.6\ex{6}$ ad	& $2.5\ex{7}$ ad	& $2.3\ex{8}$ ae	& 			& $2.5\ex{8}$ ae	&			&			\\
			&		&		&			& $\bf1.4\ex{3}$	& $\bf1.4\ex{4}$	& $\bf1.3\ex{5}$	& 			& $\bf1.4\ex{5}$	&			&			\\
BRI 0952\s		& QSO		& 4.43	& 360 e		&			&			&			&			& $4.3\ex{7}$ af	&			&			\\
			&		&		&			&			&			&			&			& $\bf1.5\ex{5}$	&			&			\\
B3 J2330		& HzRG	& 3.09	& 1950 e		&			&			&			& $1.1\ex{8}$ ag	&			&			&			\\
			&		&		&			&			&			&			& $\bf5.6\ex{4}$	&			&			&			\\
TN J0121		& HzRG	& 3.52	& 1050 e		&			&			&			& $1.3\ex{8}$ ah	&			&			&			\\
			&		&		&			&			&			&			& $\bf1.2\ex{5}$	&			&			&			\\
6C 1909		& HzRG	& 3.54	& 1470 e		&			&			&			& $1.7\ex{8}$ ai	&			&			&			\\
			&		&		&			&			&			&			& $\bf1.2\ex{5}$	&			&			&			\\
\hline 
\end{tabular}
\end{center}
$\star$ lensed source.\\
References: (a) \citet{kneib04}, (b) \citet{weiss05b}, (c) \citet{sheth04}, (d) \citet{kneib05}, (e) \citet{solomonvdb05}, (f) \citet{greve05}, (g) \citet{hainline06}, (h) \citet{tacconi06}, (i) \citet{kovacs06}, (j) \citet{neri03}, (k) \citet{genzel03}, (l) \citet{frayer99}, (m) \citet{downessolomon03}, (n) \citet{takata06}, (o) \citet{greve03}, (p) \citet{andreani00}, (q) \citet{frayer08}, (r) \citet{baker04}, (s) \citet{planesas99}, (t) \citet{brown91}, (u) \citet{solomon92a}, (v) \citet{solomon92b}, (w) \citet{downes95}, (x) \citet{barvainis94}, (y) \citet{wilner95}, (z) \citet{barvainis97}, (aa) \citet{weiss03}, (ab) \citet{hainline04}, (ac) \citet{barvainis02}, (ad) \citet{carilli02b}, (ae) \citet{cox02}, (af) \citet{guilloteau99}, (ag) \citet{debreuck03a}, (ah) \citet{debreuck03b}, (ai) \citet{papadopoulos00}.
\end{minipage}
\end{table*}

\section{Star formation model}\label{sec:sf}
Following the same strategy as in our previous work on the infrared sources \citep{righi08}, we model the star formation in a merger episode and characterize the distribution of the merging using the EPS formalism \citep{lc93}. In a very recent paper, \citet{nd08} improve the EPS formalism, but their new estimates of the merger rates only deviates significantly from the results of Lacey \& Cole for the minor mergers. For the merging mass ratios considered in our approach (0.1-10), the discrepancy in the two models is only 20\%. In the following we briefly outline the main equation of our star formation model.

In each merging episode, a given amount of gas is converted into stars. Considering two halos of mass $M_1$ and $M_2$, merging to yield a halo of mass $M=M_1+M_2$, a given amount of the halo mass is converted into stellar mass
\begin{equation}
M_{\star}^1=\frac{\Omega_{\mathrm{b}}}{\Omega_{\mathrm{m}}}\,\eta\,M_1\frac{M_2}{M/2}.
\end{equation}
The amount of stellar mass formed from the halo $M_1$ is therefore proportional to its mass and to the mass of the merging halo $M_2$, if this is sufficiently massive. In the same way, for the second halo,
\begin{equation}
M_{\star}^2=\frac{\Omega_{\mathrm{b}}}{\Omega_{\mathrm{m}}}\,\eta\,M_2\frac{M_1}{M/2}.
\end{equation}
The total stellar mass produced in this merging episode is then the sum
\begin{equation}
M_{\star}=M_{\star}^1+M_{\star}^2=4\,\frac{\Omega_{\mathrm{b}}}{\Omega_{\mathrm{m}}}\,\eta\,\frac{M_1\cdot M_2}{M}.
\end{equation}
The parameter $\eta$ is the star formation efficiency. We take this to be $5\%$, in order to match the observations of the cosmic star formation history \citep{hopkins06}.

The lifetime of the starburst phase is then introduced following the results of the several numerical simulations available in literature \citep[e.g.][]{MH94,MH96,SH05,ROB06}. There are two star formation bursts, corresponding to the first close passage of one galaxy around the other and to the final coalescence, with a typical duration of $\sim300$ and $\la100$~Myr, respectively. A third phase, identified as the star formation activity already present in the two galaxy before the interaction, is not included in the model, since it is not directly trigger by the merging. This simple parametrization allows us to derive a statistical description of the merging events as a function of their star formation rates. 

Comparing the typical timescale of the bright phase ($\sim100$ Myr) with the characteristic evolution time for old stars, which is about 1~Myr, one can easily prove that these objects are rapidly enriched with metals already at high redshift. This is the main argument that motivates the work presented in this paper.

The next step is then to connect the star formation rate with the luminosity in a given band. In our previous work, we were interested in describing a population of far-infrared sources, so we used the Kennicutt relation for that band \citep{kennicutt98}. Here we aim to describe the luminosity in the line for a set of molecular and atomic transitions that occur in the observing band of the CMB experiments. We assume that line luminosity scales with the star formation rate
\begin{equation}\label{mdotlrel}
L_{\mathrm{line}}[L_{\odot}]=\mathcal{R}\cdot \dot{M}_{\star}\,[M_{\odot}/\mbox{yr}].
\end{equation}
The constant $\mathcal{R}$ will be calibrated in the next section, taking a sample of objects of different nature into account both at low and high redshifts, along with their observed value of the luminosity in each line.

\section{Line luminosities from a sample of objects}
In the previous section we introduced a simple relation between the star formation rate and the luminosity in a given line, similar to the existing relations in different bands \citep[see][for a review]{kennicutt98}. The constant factor connecting the two quantities depends on several parameters, and it is not easy to determine it. To get a rough estimate of its value for different lines, we consider a sample of objects and retrieve data on their line luminosity from the literature.\\
For the CO transitions, we consider both a low- and a high-redshift sample. The first is derived from the list of \citet{baan08}, which contains more than a hundred sources, observed with different instruments. We selected in this list only the galaxies with optical size smaller than the beam size of the observation. This is to avoid an underestimate of the flux due to incomplete coverage of the source. More detailed studies have been done on well-known local objects like M82  \citep{weiss05a} and NGC4038/39 \citep[also known as The Antennae,][]{gao01,bayet06}. These are, however, extended sources so one has to be very careful with the observational results, since they depend on the actual region covered by the instrument. In the case of M82, \citet{weiss05a} present an accurate model of the CO emission from this galaxy, which takes both the central and the outer regions into account. The combined emission is a reliable estimate of what would be observed if M82 was shifted to cosmological distances, therefore we will use it to calibrate the $\mathcal{R}$ ratio in our model. The relatively low mass of M82 \citep[$M\sim10^{10}\,M_{\odot}$,][]{mayya06} is typical of the objects that contribute to the bulk of the correlation term estimated here ($M=10^9-10^{11}\,M_{\odot}$ for redshift $z=3-5$). The data of the low-redshift sample are presented in Table~\ref{tab:COlowz}: the star formation rates are obtained applying the \citet{kennicutt98} relation to the far-infrared ($8-1000\,\mu$m) luminosity given in the \emph{IRAS revised bright galaxies sample} \citep{sanders03}. Distances are from the same catalog: conversion from proper to luminosity distance has been computed using the cosmology-corrected redshift in the Nasa Extragalactic Database (NED). For our Galaxy we use the value of star formation rate given by \citet{cox00}. 

In the same table, we also present the ratio between dust and line luminosities in M82, using the spectrum given in \citet{lagache05} and the model of \citet{weiss05a}. In our previous paper \citep{righi08}, we demonstrated that dust emission from merging galaxies is the most important high-frequency extragalactic foreground, therefore it is very important to compare with it. As we can see from the corresponding line in Table~\ref{tab:COlowz}, the ratio of dust-to-CO-line emission grows with frequency; therefore, the first CO transitions are more effective and should be favored from the point of view of their ratio to the dust continuum.

The high-redshift sample contains many sub-millimeter galaxies. These are very massive and luminous systems, powered by intense starburst and/or AGN, and they emit the bulk of their radiation in the infrared band. This activity is triggered by merging episodes that provide large amount of gas and results in high values of the star formation rate. We retrieved from the literature the line luminosity for several objects, including QSOs and radiogalaxies. These are reported in Table~\ref{tab:COhighz}: luminosities of the lensed sources (marked with $\star$) are corrected with the magnification factors given in \citet{greve05} and \citet{solomonvdb05}. Where more than one observation were available we assumed an average between them. These observations were carried out with interferometric technique; therefore, they are more sensitive to bright small spots and not to the continuum emission. Comparing the average $\mathcal{R}$ of this sample with the one obtained for M82, we see that there is no big difference, though the CO spectral energy distribution for the high-redshift objects tends to peak at higher values of the rotational quantum number $J$. This may derive from differences in the temperature and in the density of the gas. Furthermore, in the high redshift objects, the presence of the non-negligible radiation field of the CMB contributes to increasing the excitation level of the CO molecule.

We summarize all these observational data in Fig.~\ref{fig:COplot}, where we plot the $\mathcal{R}=L/\dot{M}$ ratio as a function of the upper rotational quantum number of the transition. A higher value of $\mathcal{R}$ corresponds to a higher luminosity of the galaxy per unit star formation rate, hence to a higher amplitude in the fluctuations. We present the two samples of objects described above: local objects, low-redshift IRAS galaxies, high-redshift sub-millimeter galaxies, QSO and radiogalaxies. The spectral energy distribution of M82, for which we have the full set of lines, is overplotted. We show here also the values for our Galaxy, as observed by COBE/FIRAS \citep{wright91}, assuming a star formation rate of 3~$M_{\odot}$/yr \citep{cox00}. From this dataset we extract an estimate of the lower and upper value of $\mathcal{R}$ in every transition. Notice that local merging objects, which are intrinsically weaker than the QSO, are close to the upper limit, meaning that they have a higher luminosity per unit star formation rate. M82 and the Antennae, for example, are extremely bright in the first two transitions, compared with the other objects. There is rather large scatter in the distribution of these points at different $J$. This might be due to the uncertainties in the CO properties or more probably in the value of $\dot{M}$ measured in these objects.

The observations of bright distant quasars and galaxies show that they often have carbon and oxygen abundance close to or even higher than the solar value \citep{fan}. It is clear that to the average abundance of metals is much lower in the high-redshift universe than in the regions with intense star formation, where it can increase extremely rapidly. The Magellanic clouds (LMC and SMC) have significantly lower chemical abundance than our Galaxy; nevertheless, the luminosities of the CO lines are comparable \citep{cohen88,ck96}. This problem is of particular interest for the first line of the CO, since it is heavily saturated and therefore its brightness depends weakly on chemical abundance. On the other hand, higher-$J$ lines are usually optically thin and therefore their brightness strongly depends on the chemical abundance. This dependence might be non-linear, with an index close to $1.5-2$, leading to lower CO emission in metal-poor systems. This is also connected with the low dust abundance and a consequently less efficient shielding from the ultraviolet radiation, which destroys CO molecules (Combes, priv. comm.). However, in this paper we are dealing with star forming regions, in which rapidly evolving massive stars efficiently enrich the medium with significant amounts of metals. M82, which is the object used to calibrate most of the results presented here, has a metallicity 1.5 times the solar value \citep{galliano08}.\\
The saturation of the first two CO lines might lead to a non-linear dependence of the line luminosity with the star formation rate. This was shown by \citet{narayanan07}: using a radiative transfer model combined with numerical simulations they found that this dependence is linear for the higher transitions ($J>3$), but might become non linear at lower $J$.

\begin{figure}
  \resizebox{\hsize}{!}{\includegraphics{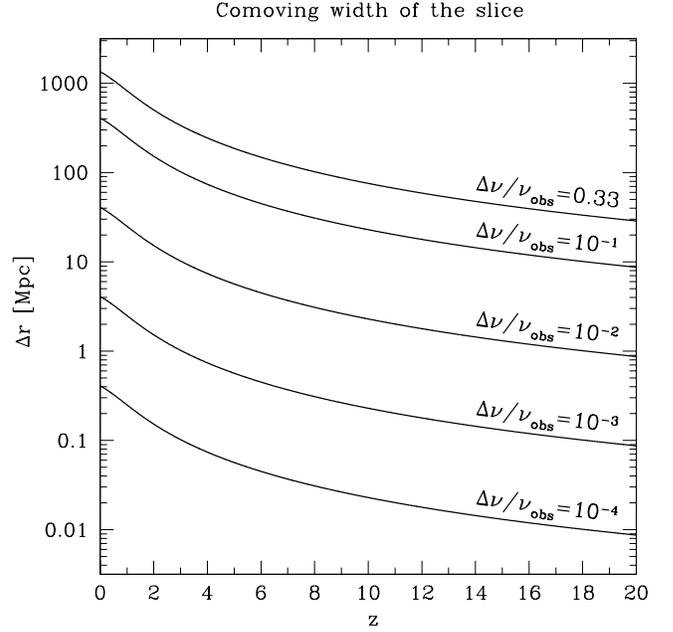}}
  \caption{The actual comoving width of the slice (in Mpc) as a function of redshift, probed by different values of the spectral resolution $\Delta\nu/\nu_{\mathrm{obs}}$.}
  \label{fig:deltar}
\end{figure}

\begin{figure*}
  \centering
  \includegraphics[width=5.9cm]{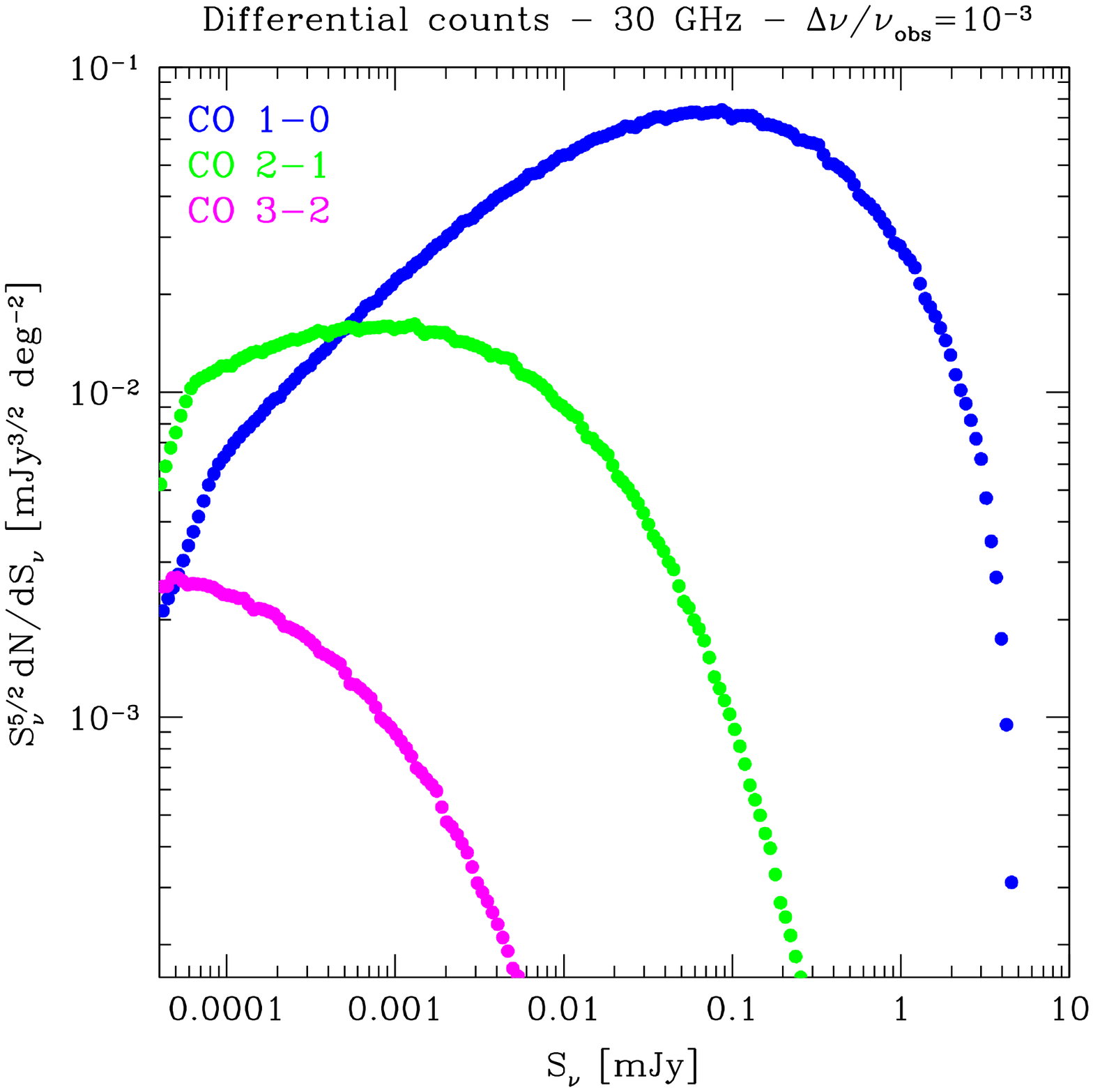}
  \includegraphics[width=5.9cm]{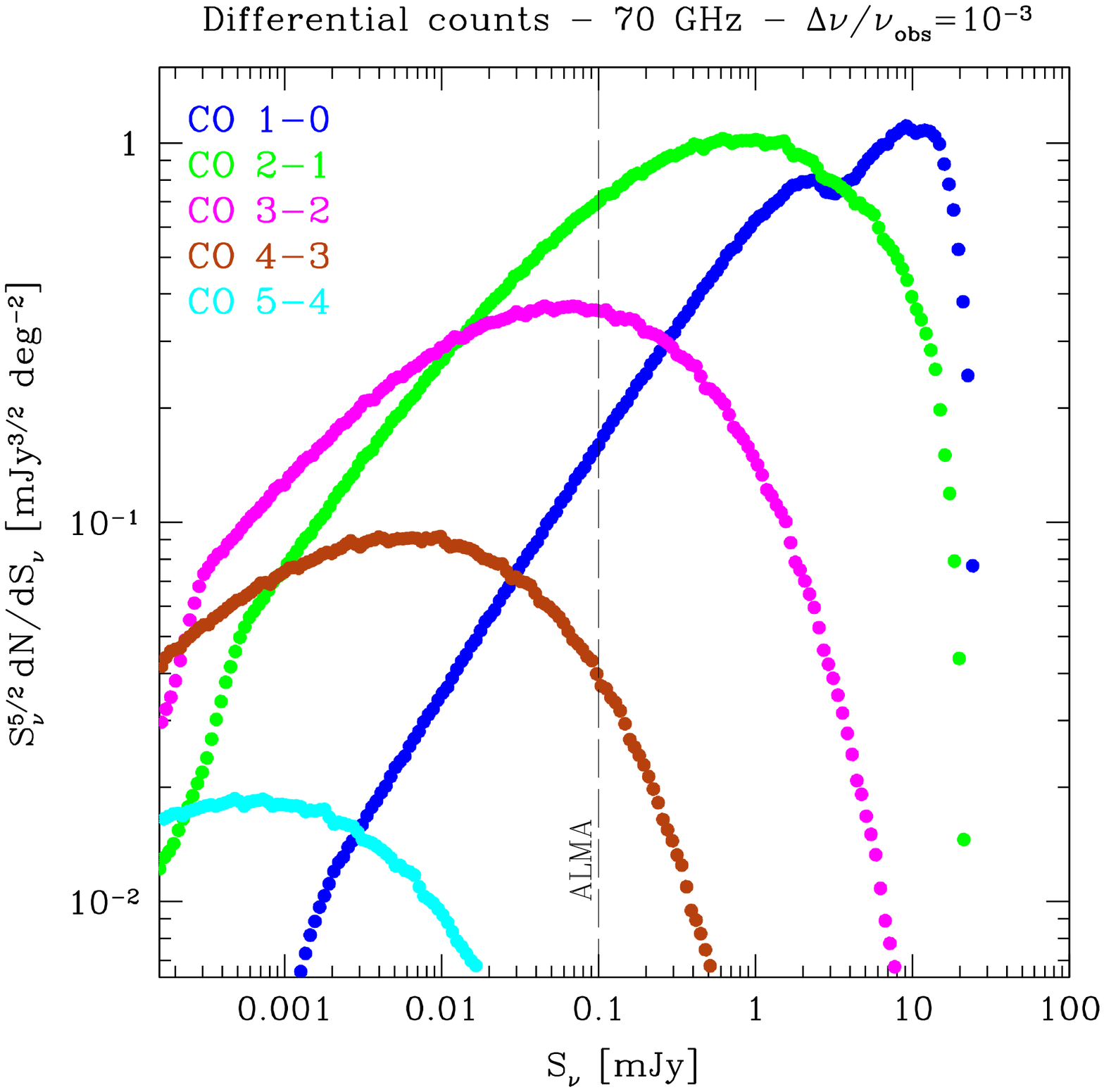}
  \includegraphics[width=5.9cm]{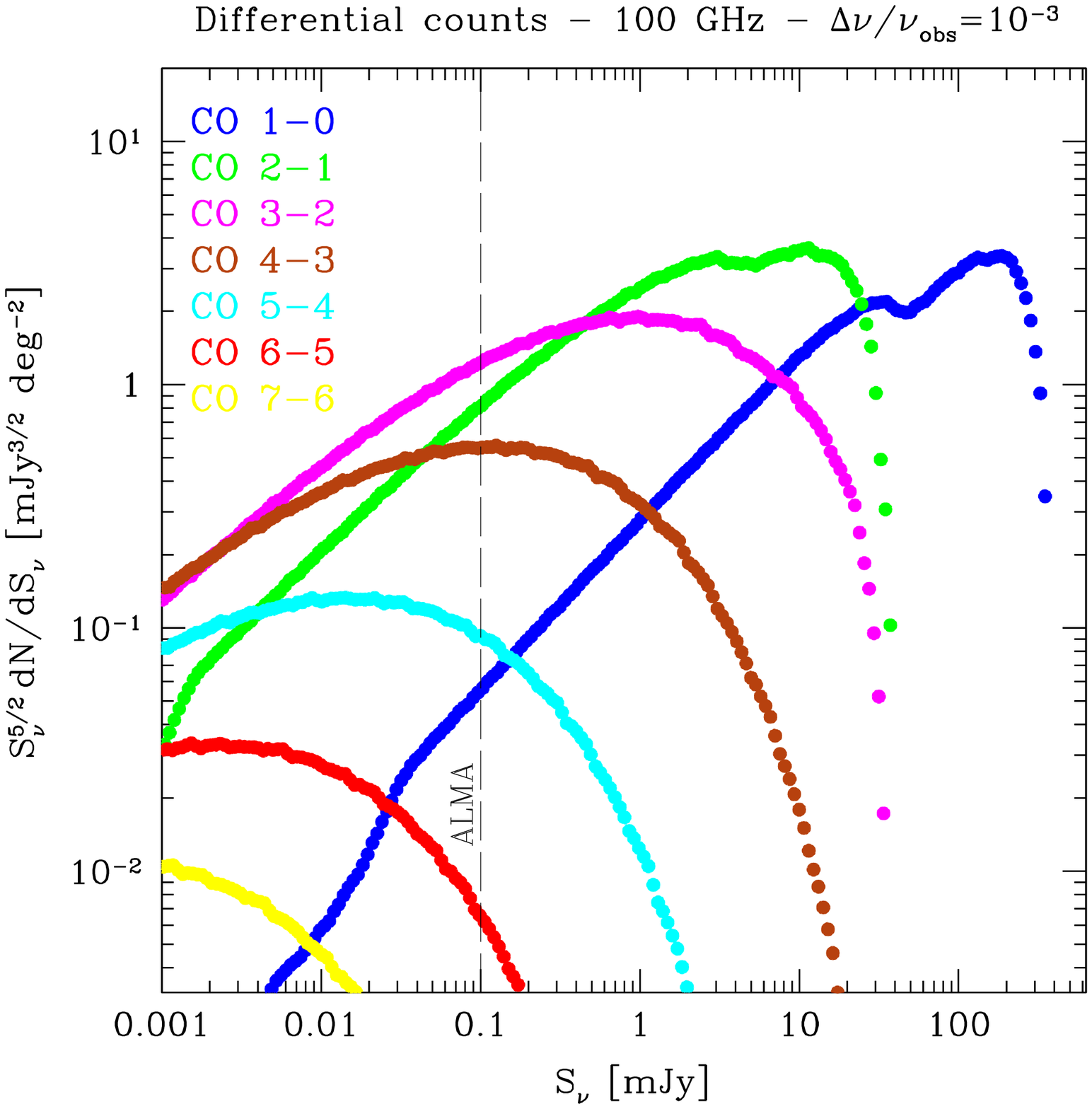}
  \caption{The predicted differential source counts for the CO lines at 30, 70, and 100~GHz, for a spectral resolution $\Delta\nu/\nu_{\mathrm{obs}}=10^{-3}$. The vertical line is the expected sensitivity of ALMA to the line emission, computed for the same spectral resolution and an integration time of 3 hours.}
  \label{fig:counts_CO}
\end{figure*}


\section{The angular power spectrum}\label{sec:powspec}
In the Appendix~\ref{sec:lineformal} we present a detailed derivation of the expressions we used for the angular power spectrum generated by the emission in these lines. Here we briefly outline the main results. First of all, we should note that the measured signal ${\tilde I}_{\nu}(\vnh)$ is actually a convolution of the underlying signal $I_{\nu} (\vr = r\vnh )$ with an experimental window function ${\cal B}(\vr )$:
\begin{equation}\label{eq:conf1}
{\tilde I}_{\nu} (\vr) = \int d\vr' {\cal B}(\vr-\vr') \, I_{\nu}(\vr' ).
\end{equation}
This window function ${\mathcal B}$ simply accounts for the finite angular and spectral resolutions of the observing device. The variance of the signal can be expressed as a function of the measured power spectrum ($|{\tilde I}_{\nu, \vk}|^2$)
\begin{equation}\label{eq:ps1}
\langle {\tilde I}_{\nu}^2 \rangle=\int \dkthree \, |{\tilde I}_{\nu,\vk}|^2=\,\int \dkthree  \,|{\cal B}_{\vk}|^2 \,P_s(k)\,|\bar{I}_{\nu}|^2,
\end{equation}
where $ \bar{I}_{\nu}$ denotes the volume weighted intensity amplitude for a given source, and $P_s(k)$ is the power spectrum describing the spatial
distribution of the sources. This integral reflects the contribution of anisotropy (given by the power spectrum $P_s(k)$) to each $k$ scale range,
and this depends on the size and distribution of the sources.  If these are randomly (Poisson) distributed in space, then $P_s(k) = 1 / n$, with $n$ the
average source number density. This means that {\em all} scales contribute with the same amount of anisotropy (as given by the $k$-independent power
spectrum $P_s$), down to a minimal linear scale (maximum $k$) corresponding to either the size of the source or the resolution element of the instrument.  If the source size is taken as arbitrarily small, and we fix the angular resolution, then the variance or measured anisotropy will increase if the instrument is sensitive to smaller and smaller scales {\em in the radial direction}; i.e., it will scale as $k_{\max}\sim 1/(\Delta z)\sim 1/(\Delta \nu)_{\mathrm{inst}}$, with $z$ denoting the linear scale along the radial direction. More generally, if sources are Poisson distributed, any improvement in the spectral/angular resolution that makes the experiment sensitive to smaller scales (that are still larger that the typical source size) will yield an increase in the measured power.

\begin{figure}
  \resizebox{\hsize}{!}{\includegraphics{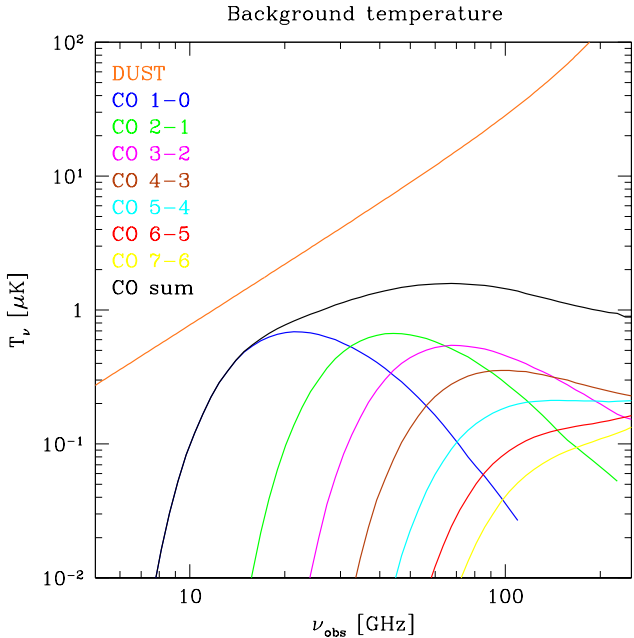}}
  \caption{The contribution to the cosmic microwave background radiation (in temperature units) from the different CO lines and of their sum (black line), compared with the signal from dust (orange).}
  \label{fig:background}
\end{figure}

However, we know that, in the universe, sources tend to be clustered in regions of $L_c\sim$ 15-25 $h^{-1}$ Mpc (comoving size) that eventually will become
superclusters of galaxies. The distribution of these regions will introduce more anisotropy, but only on scales that are actually larger
than $L_c$. If the frequency/angular resolution of the experiment is at some point able to resolve this scale, then further improving such resolution will make no difference. This can be rephrased in Fourier space as follows. If sources are distributed such that their power spectrum is
proportional to that of the underlying density field ($P_s(k) \propto P_m(k)$), then for some large $k$ the contribution of the power spectrum to
the integral of Eq.~(\ref{eq:ps1}) will be negligible (since for large $k$ we have that $P_m(k)\rightarrow 0$). In this scenario, increasing the spectral
resolution will not change the integral over $P_s(k)$. These two distinct regimes will be addressed in detail below.

After projecting the power spectrum on the sphere, we find expressions for the Poisson ($C_l^P$) and correlation ($C_l^C$) angular power spectra (see
Appendix). We recall that their relation to the angular correlation function reads as
\begin{equation}\label{eq:cforig}
\langle {\tilde I}_{\nu} (\vnh_1)  {\tilde I}_{\nu} (\vnh_2) \rangle = \sum_l\frac{2l+1}{4\pi}\left(C_l^P+C_l^C\right)P_l(\vnh_1
\cdot \vnh_2),
\end{equation}
with $P_l(\vnh_1 \cdot \vnh_2)$ the Legendre polynomia of order $l$. In this approach the resolution is expressed in terms of a comoving width $\Delta r$ probed at a given redshift by a certain $\Delta\nu/\nu_{\mathrm{obs}}$. It is straightforward to derive the relation
\begin{equation}
\Delta r = \frac{\Delta r}{\Delta z}\,\frac{\Delta z}{\Delta\nu}\,\Delta\nu=\frac{cH_0^{-1}}{E(z)}\,\frac{\Delta\nu}{\nu_{\mathrm{obs}}},
\end{equation}
where $\Delta\nu/\nu_{\mathrm{obs}}$ refers to the observing frequency and $E(z)=[\Omega_{\mathrm{m}}(1+z)^3+\Omega_{\Lambda}]^{1/2}$. From this equation, it is clear that, for a fixed spectral resolution, the actual comoving width decreases with increasing redshift, as it is shown in Fig.~\ref{fig:deltar}.

\section{Results for the CO lines}\label{sec:res}

\begin{figure*}
  \centering
  \includegraphics[width=5.9cm]{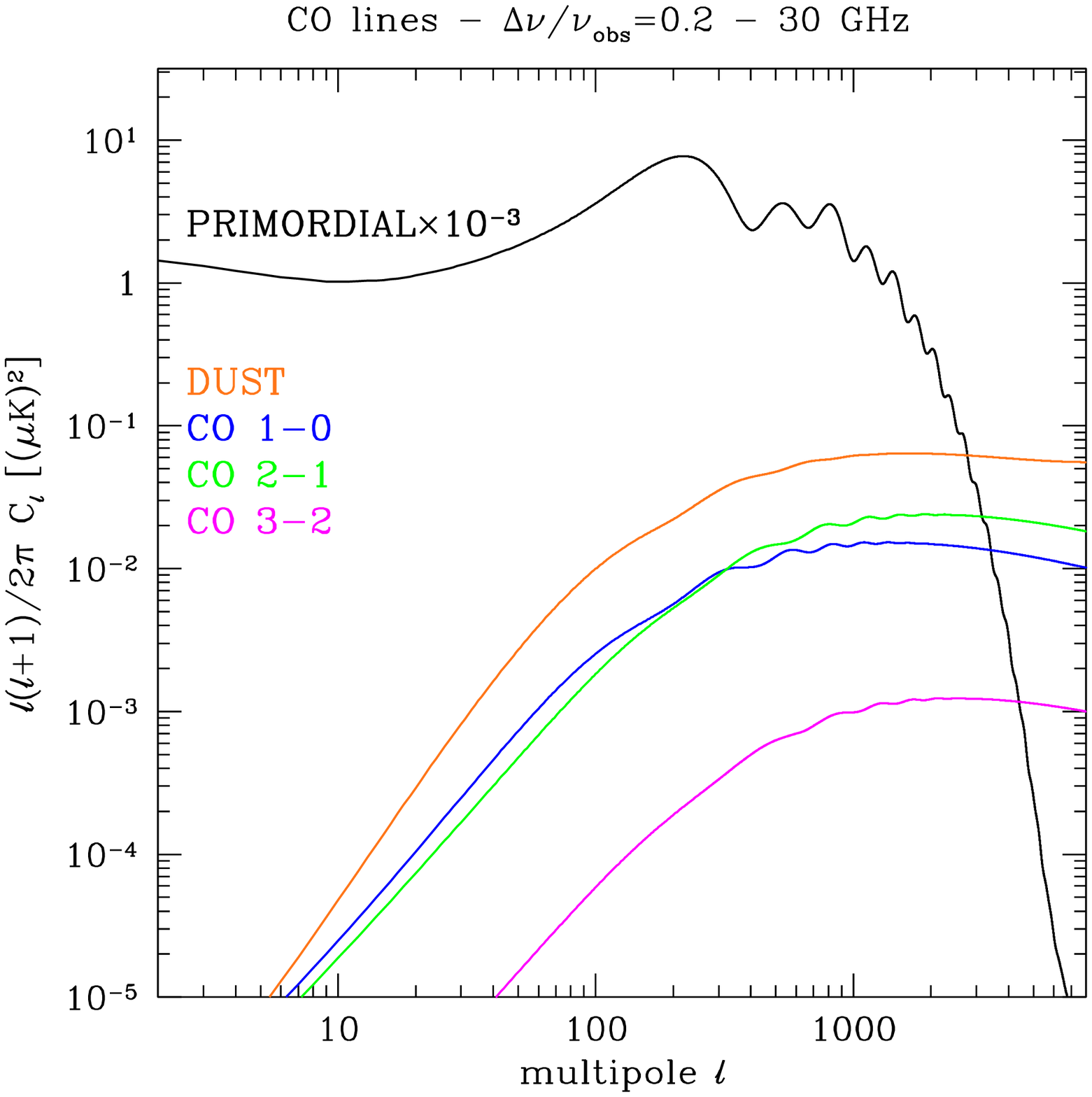}
  \includegraphics[width=5.9cm]{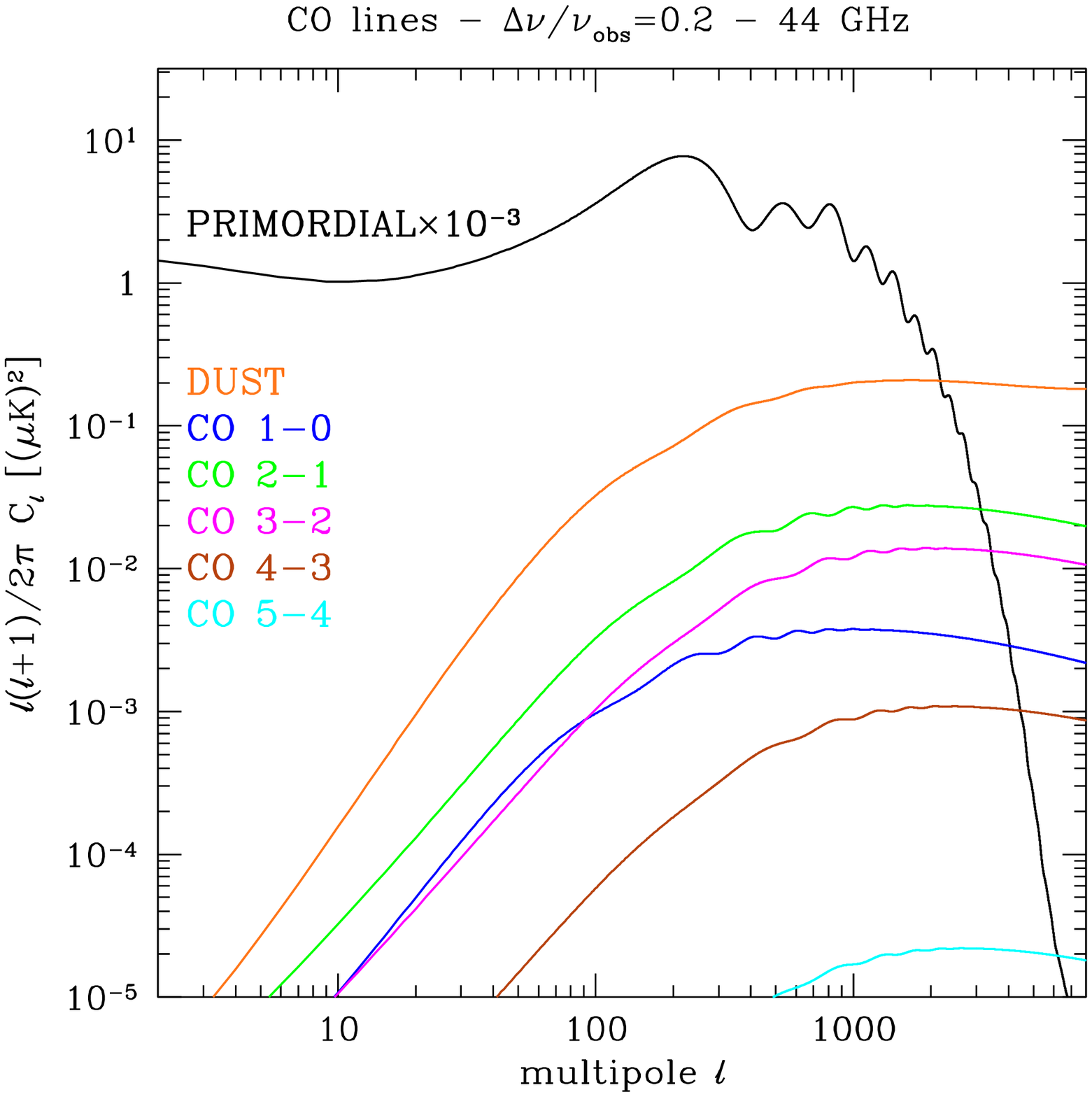}
  \includegraphics[width=5.9cm]{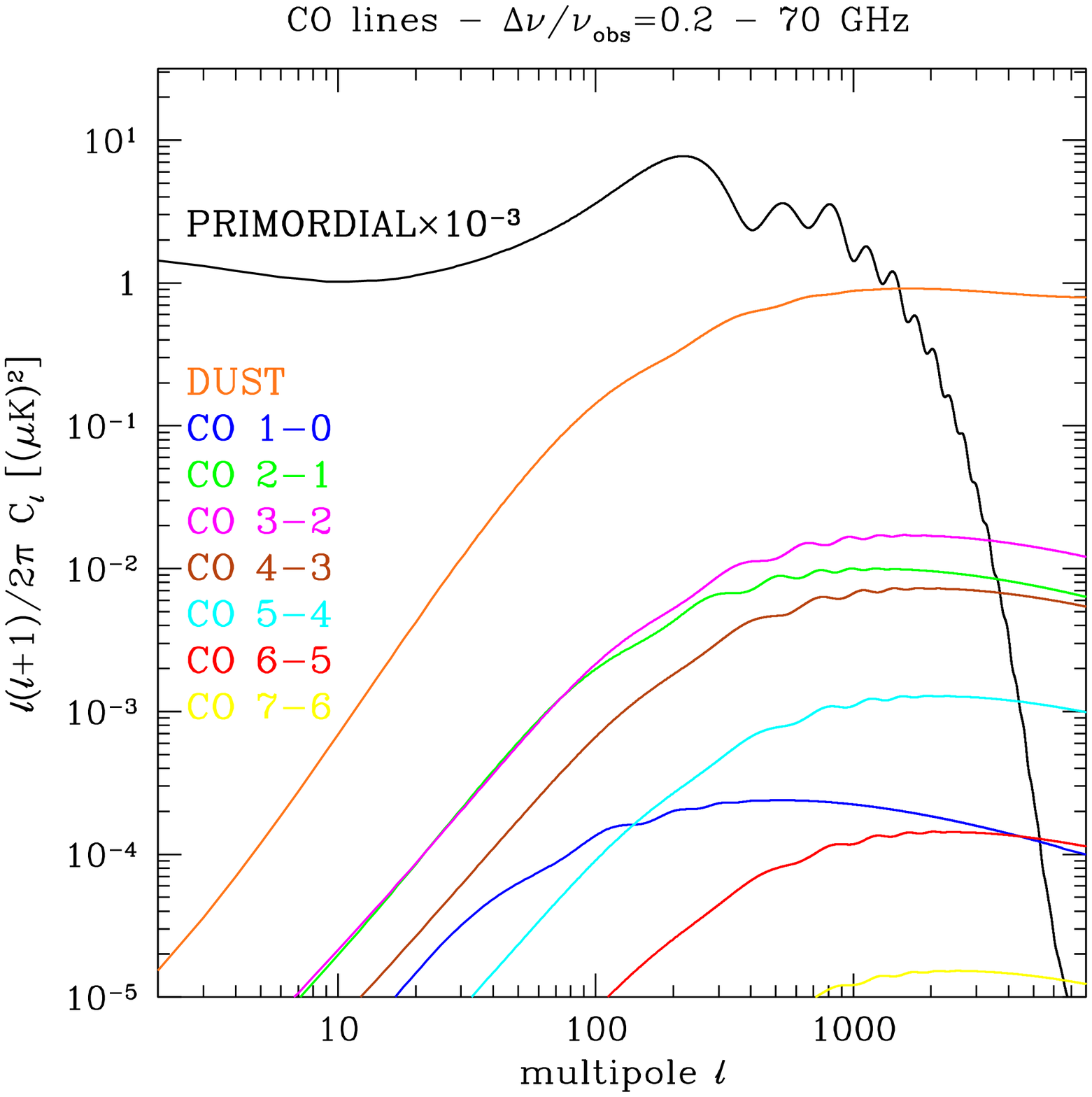}
  \caption{The correlation signal for the CO emission lines at the LFI frequencies and for a spectral resolution $\Delta\nu/\nu_{\mathrm{obs}}=0.2$, using the $\mathcal{R}$ ratio from M82. The black line is the primordial signal of the CMB divided by 1000, the orange line is the signal from dusty merging star-forming galaxies. Other colors identifies the lines as indicated by the labels.}
  \label{fig:cl_CO}
\end{figure*}

\begin{figure}
  \resizebox{\hsize}{!}{\includegraphics{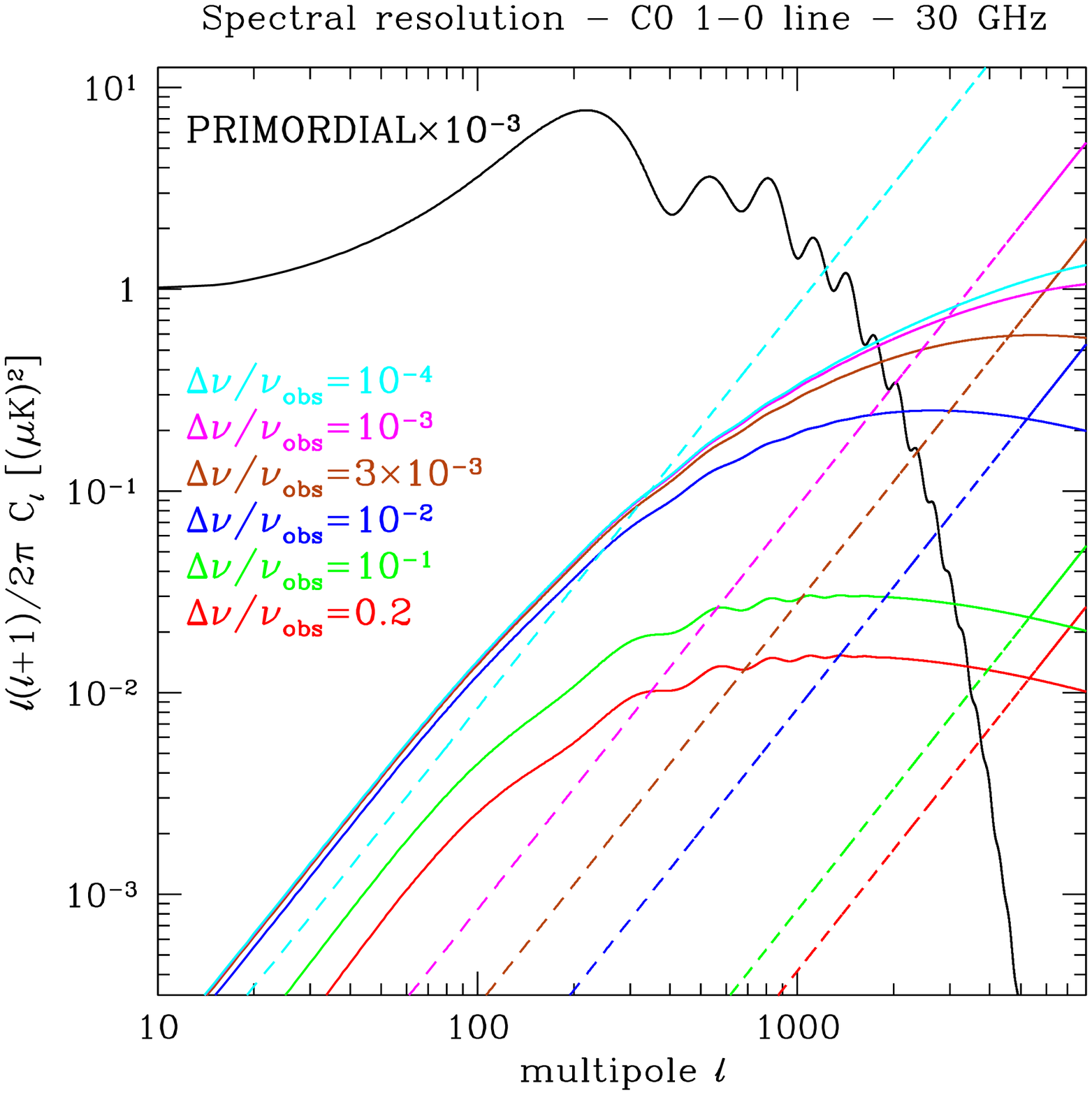}}
  \caption{The change in amplitude of the correlation (solid line) and Poisson (dashed line) signal for the CO~(1-0) line for different values of the spectral resolution $\Delta\nu/\nu_{\mathrm{obs}}$, at 30~GHz. The red and cyan lines are almost superimposed, meaning that, for these resolutions, the signal has already reached a constant value. The Poisson term, on the other hand, grows linearly with decreasing values of the spectral resolution. The primordial signal (black solid line) has been divided by 1000.}
  \label{fig:clresol}
\end{figure}

\begin{figure}
  \resizebox{\hsize}{!}{\includegraphics{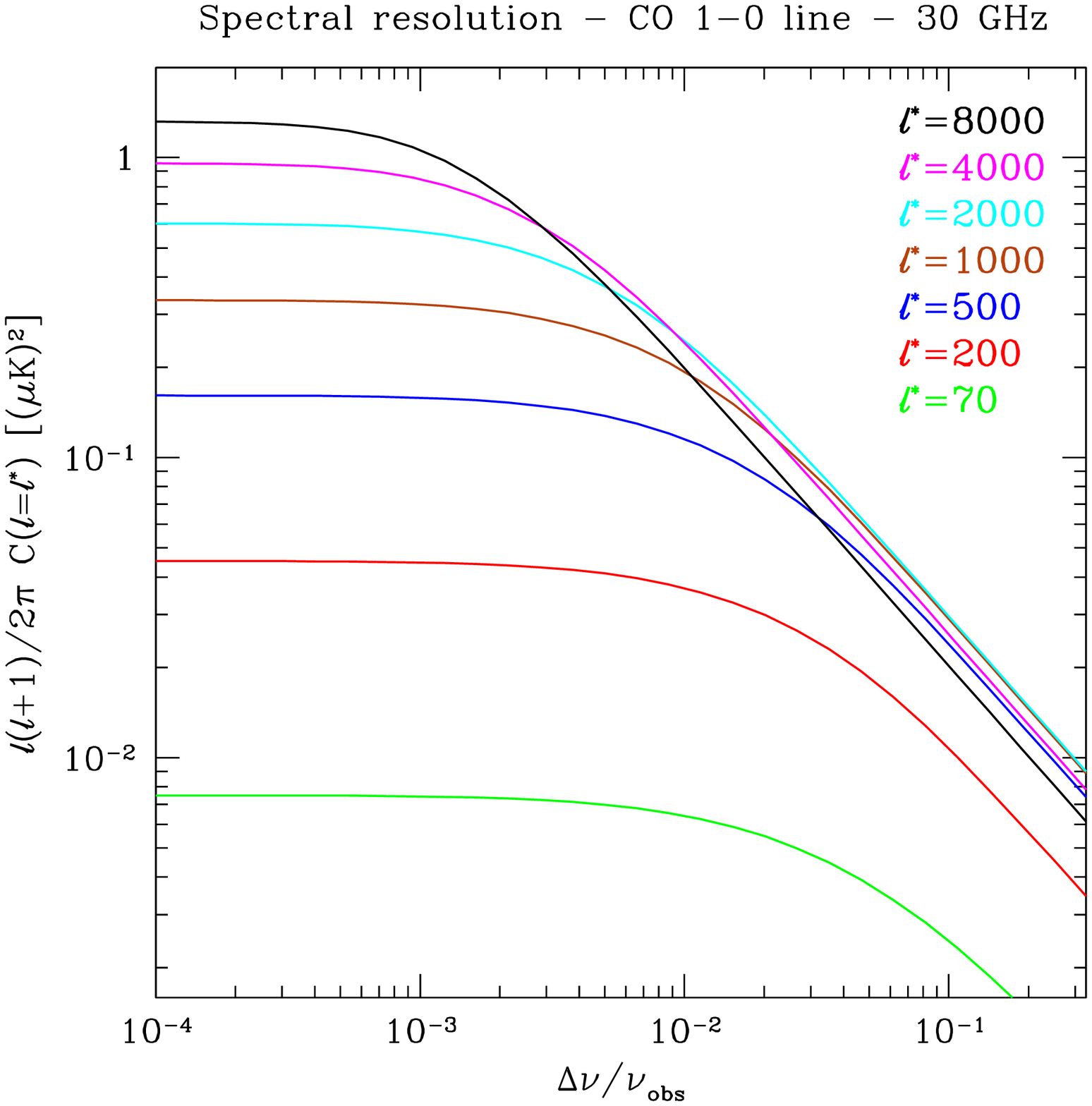}}
  \caption{The dependence of the amplitude of the correlation signal on the spectral resolution of the observing instruments, for the CO~(1-0) line at 30~GHz and for several value of the multipole index $l$.}
  \label{fig:specdep}
\end{figure}

\begin{figure}
  \resizebox{\hsize}{!}{\includegraphics{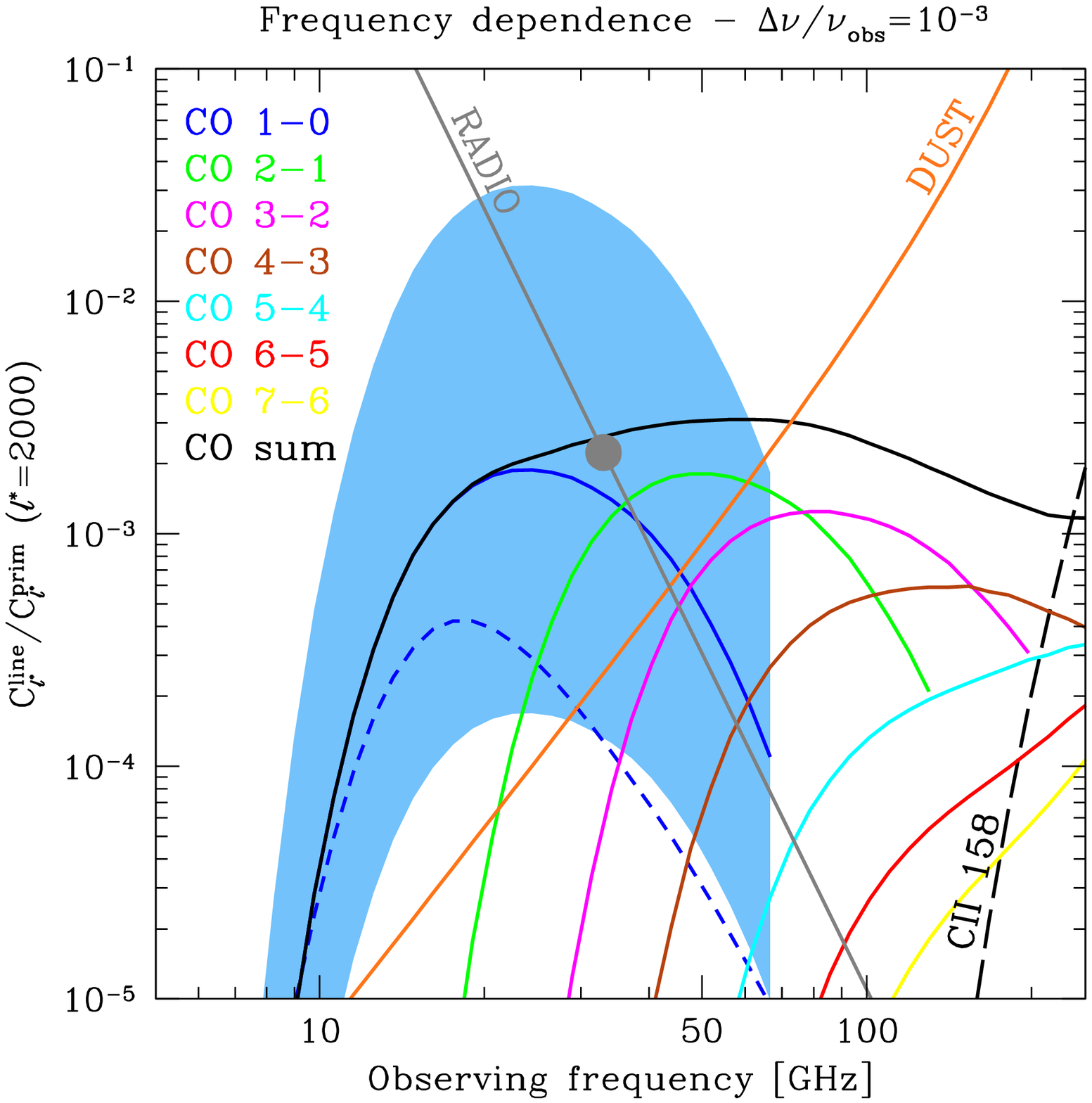}}
  \caption{The ratio of the correlation signal from CO emission to the primordial CMB signal at $l=2000$ as a function of frequency and for a spectral resolution $\Delta\nu/\nu_{\rm obs}=10^{-3}$. The black solid line is the sum of the correlation from the first seven CO lines, computed by assuming the $\mathcal{R}$ ratio for M82. The blue short-dashed line is obtained with the model based on simple Press-Schechter distribution. The black long-dashed line is the signal from CII 158~$\mu$m line. The contribution from dust emission in merging galaxies is shown with the orange line. The gray point is the Poisson at 33~GHz from radio sources, estimated from the \citet{dezotti05} model, assuming a cutoff of $0.1$~mJy. The frequency dependence of this signal (gray line) has been computed assuming $S_{\nu}\propto\nu^{-\alpha}$ with $\alpha=0.4$ \citep{toffolatti05}. The shaded region around the curve for the first CO line represents the range of uncertainties in the correlation signal, according to the different values of $\mathcal{R}$ for the sample of objects in Fig.~\ref{fig:COplot} and Tables~\ref{tab:COlowz}~and~\ref{tab:COhighz}. The first CO line is especially suitable for observations of higher redshifts because, at higher frequencies, we simultaneously observe the contributions from two or three slices of the universe.}
  \label{fig:freqdep_nu-3}
\end{figure}

\subsection{Source counts and background intensity}
In Fig.~\ref{fig:counts_CO} we present the differential source counts in the CO lines, predicted by our model at 30, 70, and 100~GHz, for a spectral resolution $\Delta\nu/\nu_{\mathrm{obs}}=10^{-3}$ and using M82 to calibrate the $\mathcal{R}$ ratio. The vertical lines in each plot show the sensitivity of ALMA to line emission \citep[$5\sigma$ value, for 3 hours integration.][]{ALMA}: this value is approximately constant at these frequencies. The flux in the line is computed as
\begin{eqnarray}
S_{\nu}&=&\frac{\displaystyle\int d\nu_{\mathrm{obs}}\,\phi_{\mathrm{instr}}\displaystyle\frac{L_{\nu}}{4\pi r^2(1+z)}}{\int d\nu_{\mathrm{obs}}\phi_{\mathrm{instr}}}=
\displaystyle\int d\nu_{\mathrm{obs}}\,\phi_{\mathrm{instr}}\frac{L_{\mathrm{bol}}\phi_{\mathrm{th},\nu}}{4\pi r^2(1+z)}=\nonumber\\
&=&\frac{1}{4\pi r^2(1+z)}\frac{L_{\mathrm{bol}}}{(\Delta\nu)_{\mathrm{instr}}}=\frac{\tilde{L}_{\nu}}{4\pi r^2 }\left(\frac{\Delta\nu}{\nu_{\mathrm{obs}}}\right)^{-1},
\end{eqnarray}
where $\tilde{L}_{\nu}=L_{\mathrm{bol}}/\nu$, $L_{\mathrm{bol}}$ is the bolometric luminosity of the line and $r$ the comoving distance. This expression shows the explicit dependence of the measured flux on the spectral resolution of the instrument. The amount of sources at a given frequency is also dependent on the spectral resolution, just because increasing the value of $\Delta\nu/\nu_{\mathrm{obs}}$ corresponds to probing larger volumes of the universe.

Given the source number counts, we can compute the contribution of each line to the cosmic microwave background temperature (Fig.~\ref{fig:background}). The total emission of the CO lines is practically constant with frequency in  the very broad $20-200$~GHz range. Its value is close to 1~$\mu$K, and it is thousand times weaker than the present day uncertainties in the temperature of monopole according to COBE/FIRAS results \citep{fixsenmather02}. The first CO line peaks at frequency $\sim15-20$~GHz, corresponding to an emission redshift $z\simeq5-7$. The dust emission from merging galaxies contributes two or three times more at frequencies around $15-20$~GHz, increasing then to a value of $100$~$\mu$K at 200~GHz.

At 30~GHz, most of the contribution to the background in the first CO transition is due to low-flux sources ($S_{\nu}=10^{-4}-10^{-2}$ mJy), which cannot be detected directly. But since their contribution to the correlated angular fluctuations is high, observations of the $C_l$'s with different spectral resolution provide us with a powerful tool for studying the properties of the high-redshift population of CO-emitting galaxies and its evolution in time. We discuss this in the next section.

\subsection{Power spectrum of angular fluctuations}
In this section we present the angular power spectra arising from the line emission of CO in the star-forming regions. Such a power spectrum is the sum of a Poisson term $C_l^P$ and a correlation term $C_l^C$.

The emission lines from carbon monoxide are associated with the rotational transitions between different states $J$. Their emission rest frequencies lie in the range $100-800$~GHz making them of particular interest for the low-frequency experiments. Unfortunately, there are few observations of this molecule in the sample that we have selected, so we were only able to retrieve the full data set for M82. We use this object to calibrate the $\mathcal{R}$ ratio in our computations (see Table~\ref{tab:COlowz} for details), but we show also the expected upper and lower limits in the amplitude of the signal, according to the value presented in Fig.~\ref{fig:COplot}. The correlation terms of the angular power spectrum for each CO line are shown in Fig.~\ref{fig:cl_CO} at the three frequencies of PLANCK's LFI instrument and for the corresponding spectral resolution $\Delta\nu/\nu_{\mathrm{obs}}=0.2$. As expected, the signal at low frequencies is dominated by the lines corresponding to the first two transitions ($2-1$ and $1-0$), while higher states are important only at higher frequencies. This might appear paradoxical, since the SED of CO peaks at higher $J$. One has to bear in mind, however, that the power spectra are presented in temperature units and normalized to the CMB blackbody.

Another interesting aspect is that the emission from the first two transitions of the CO might be slightly amplified  by the contribution from CN, HCN, HNC, and HCO$^+$. These molecules form in the same environment of CO and has a very similar structure. Their resonant frequencies are very close (within $\sim20\%$) to the 115 and 230~GHz transitions of CO. Their signal is then summed to the CO, with a slight shift in frequency and redshift, which leads to a smoothing effect. The luminosity of such lines in nearby objects is typically $\sim10\%$ of the CO lines \citep{baan08}.

As we have seen, in the literature there are also several observations of high-redshift sub-millimeter galaxies. These objects are thought to be the result of very massive and gas-rich mergers, with very high star formation rates, on the order of $10^3$~$M_{\odot}$/yr. According to \citet{alexander05b,alexander05a} around 40\% of the sum-millimeter galaxies (SMG) population hosts an AGN, but these only contribute to 10-20\% of the total energy output in the far-infrared \citep{pope08}. In Table~\ref{tab:COlowz} and Table~\ref{tab:COhighz} we summarize the data about CO emission in these objects collected from the literature. We note that the line luminosity in these objects is extremely high, even for the higher $J$ transitions, but also given the high value of the star formation rates, the resulting $\mathcal{R}$ does not change too much. In addition, these very luminous objects are also very rare and therefore do not contribute significantly to the correlation term, which is due mainly to the most abundant, fainter objects.

\subsubsection{Dependence on the spectral resolution}
As shown in Sec.~\ref{sec:powspec}, the spectral resolution of the instrument is extremely relevant in determining the amplitude of this signal. This is one of the most interesting results of this work: we demonstrate analytically (Appendix~\ref{sec:lineformal}) that the amplitude of the correlation term is expected to grow for lower values of $\Delta\nu/\nu_{\mathrm{obs}}$ to some threshold value, beyond which it reaches a convergence level and stays constant for even lower $\Delta\nu/\nu_{\mathrm{obs}}$-s.  This can be easily seen in Fig.~\ref{fig:clresol} for the CO~(1-0) line: the amplitude of the correlation term gains more than one order of magnitude as the relative spectral width is decreased to $10^{-3}$, but shows practically no change if this value is further reduced to $10^{-4}$. On the other hand, the Poisson term grows linearly with decreasing values of $\Delta\nu/\nu_{\mathrm{obs}}$; therefore, the relative importance of the two terms changes as a function of the spectral resolution.

We should keep in mind, however, that the actual amplitude of the Poisson term strongly depends on the ability of the observing instruments to isolate and remove the bright individual sources. The bulk of the power in the amplitude of the Poisson term, in fact, is generated by rare and bright sources. Such sources can be removed from the maps observing the same region with high enough angular resolution. This has practically no effect on the amplitude of the correlation term, which is generated by the abundant low-flux sources. This is clear from Table~\ref{tab:percentages}, where we compute the contribution to $C_l^C$ and $C_l^P$ of the CO~(1-0) line at 30~GHz from different flux decades; with a 1-mJy flux cut-off, one can drastically decrease the Poisson amplitude of $70\%$, while decreasing the correlation of only $10\%$.

\begin{table}
\caption{Contribution to the amplitude of the correlation and Poisson terms of the CO~(1-0) line from different intervals of flux. The observing frequency is 30~GHz and the spectral resolution is $10^{-3}$.}
\label{tab:percentages}
\centering
\begin{tabular}{lcc}
\hline
\hline
$S_{\nu}$ [mJy]			& $\%$ of $C_l^C$	& $\%$ of $C_l^P$ \\
\hline
$S_{\nu}<10^{-4}$			& 9			& 0			\\
$10^{-4}\le S_{\nu}<10^{-3}$	& 23			& 1			\\
$10^{-3}\le S_{\nu}<10^{-2}$	& 30			& 4			\\
$10^{-2}\le S_{\nu}<10^{-1}$	& 26			& 25			\\
$10^{-1}\le S_{\nu}<1      $	& 11			& 53			\\
$S_{\nu}>1$				& 1			& 17			\\
\hline
\end{tabular}
\end{table}

The amplitude of the correlation term as a function of the spectral resolution is explored in more detail in Fig.~\ref{fig:specdep} for the same line. Here we plot the value of the correlation term at different multipoles $l$ (in the range $l=70-8000$, corresponding to angular scales $\theta\simeq 2.5^{\circ}-1'$). The signal grows uniformly and then reaches a plateau around a value $\Delta\nu/\nu_{\mathrm{obs}}\sim10^{-3}$. The position of this convergence point depends both on the redshift of the line and on the multipole~$l$. In general larger scales (smaller $l$) and higher redshift lines seem to converge earlier, i.e. for higher value of $\Delta\nu/\nu_{\mathrm{obs}}$.

This can be explained as follows. The $C_l$'s for the correlation term can be expressed as an integral in the line-of-sight component of the Fourier mode ($k_z$)
over the halo power spectrum,
\begin{equation}
C_l^C \propto \int_0^{k_{z,max}} \frac{dk_z}{2\pi} P_h(k_{\perp}, k_z)
\left| W(k_z)\right|^2,
\label{eq:corr_cls}
\end{equation}
where $W(k_z)$ is the Fourier window function along the radial direction, related to the spectral resolution of the experiment. The symbol $k_{\perp}$ represents the Fourier model included in the plane of the sky, and it is related to the $l$ angular multipole via $k_{\perp} = l/r$, with $r$ the comoving distance to the observed redshift. The window function $W(k_z)$ is such that $k_z \rightarrow 0$, $W(k_z) \rightarrow 1$, but for $k_z \gg 2\pi/L_{\rm sr}$ (with $L_{\rm sr} = cH^{-1} \Delta \nu /\nu_{\rm instr}$ the length associated to the spectral resolution), then $W(k_z) \rightarrow 0$. In practice, the effect of this window function is therefore to change the upper limit of the integral in Eq.~(\ref{eq:corr_cls}) by $k_{\rm sr}\equiv 2\pi/L_{\rm sr}$. Let us assume that the halo power spectrum $P_h(k)$ is proportional to the linear matter power spectrum $P_m(k)$, with $k = (k_z^2 + k_{\perp}^2)^{1/2}$. For our choice of $l$ and observing frequency (or redshift), at $k_z=0$ the power spectrum is already being evaluated at $k$'s comparable to or larger than $\sim 0.01\, h^{-1}$~Mpc for which it shows its maximum. That means that, as long as $k_z << k_{\perp}$, $P_h(k) \simeq P_h(k_{\perp})$ and when $k_z$ becomes comparable to or larger than $k_{\perp}$, then $P_h(k) \rightarrow 0$, since the power spectrum drops very fast ($\propto k^3$) when probing scales that entered the horizon during the radiation-dominated era. On these scales there should be strong non-linear corrections to the power spectrum, which do not, however, significantly change the scaling of the integral in Eq.~(\ref{eq:corr_cls}). This integral can be approximated by $C_l \propto k_{\rm sr} \;P_h(k_{\perp}) \propto \nu_{\rm obs} /(\Delta \nu)\; P_h(k_{\perp})$  for {\em large} widths $\Delta \nu /\nu_{\rm obs}$, until it reaches a {\it plateau} for lower values of $\Delta \nu /\nu_{\rm obs}$. This {\it plateau} will be reached later (that is, for higher values of $k_{\rm sr}$ or lower values of the spectral width) for correspondingly higher values $k_{\perp}$ (or $l$).

Observing in narrow spectral channels increases the sensitivity and observing time requirements. In the future, however, it will be possible to combine CMB measurements carried out at similar frequencies but with different spectral resolutions. For instance, future Planck LFI's channel at 30~GHz will scan a similar spectral range to CBI \citep{padin02} or VSA \citep{watson03}, but with a broader frequency response: $\Delta \nu / \nu_{\rm obs} \simeq 0.2$ for Planck, but  $\Delta \nu / \nu_{\rm obs} \simeq 0.05$ for VSA and $\Delta \nu / \nu_{\rm obs} \simeq 0.03$ for CBI. Therefore, by studying the difference in the power spectra of the common areas, it will already be possible to establish constraints on the CO abundance. At 30~GHz we would simultaneously observe two different slices of the universe at $z\sim2.8$ and $z\sim6.7$. But at lower frequency we have the opportunity to observe a single slice (i.e. at $z\sim4.8$ for 20~GHz). Unfortunately, these limits would only be imposed up to a maximum multipole of $l\sim 500$, due to the relatively poor angular resolution of LFI's 30~GHz channel. Future ground-based experiments with arrays of detectors should be much more sensitive and with high enough spectral resolution could make such observations possible.

In Appendix~\ref{sec:method} we discuss a simple technique for separating the line signal from the continuum, using different spectral resolutions. 

\begin{figure}
  \resizebox{\hsize}{!}{\includegraphics{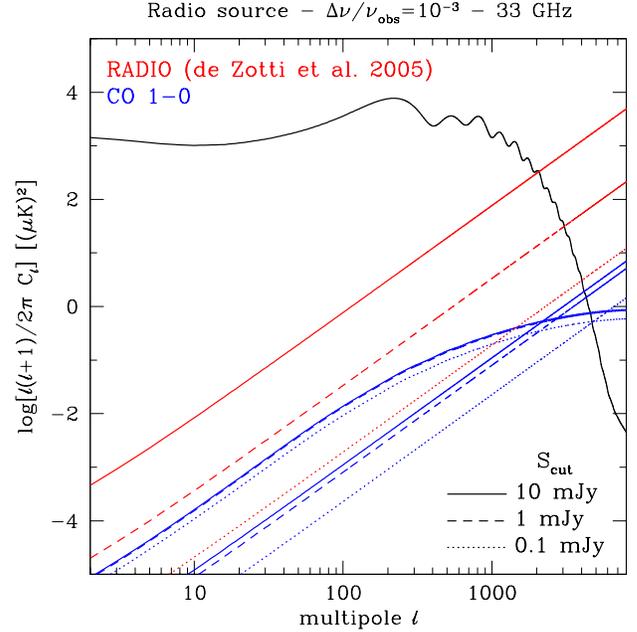}}
  \caption{The contribution of the radio sources to the power spectrum at 33~GHz (red), compared with the CO~(1-0) line (blue). Different flux cutoffs $S_{\mathrm{cut}}$ are applied: $10$~mJy (solid lines), $1$~mJy (dashed lines) and $0.1$~mJy (dotted lines). The signal due to radio sources is very sensitive to this value, while the power in the CO is much less affected, being generated by sources with fluxes much lower than $S_{\mathrm{cut}}$.}
  \label{fig:clradio}
\end{figure}

\subsubsection{Dependence on the observing channel}
In this section we explore the amplitude of the signal in different observing frequencies. We consider the range of frequency $5-300$~GHz, which will be covered by several future experiments like SKA, PLANCK, ACT, SPT, and ALMA. In Fig.~\ref{fig:freqdep_nu-3} we plot the ratio of the correlation signal to the primordial CMB signal at the fixed multipole $l=2000$: here it is very clear how the low-frequency range ($\nu\la100$~GHz) is dominated by the CO emission, while at a higher frequency, the signal from dust is much stronger and dominates the signal from the CII 158~$\mu$m line. The sum from the CO lines is computed by assuming that the cross-terms are negligible. This shows that in the frequency range $30-60$~GHz, the contribution from the first two CO lines might be the most important extragalactic foreground for the observations of the primordial angular fluctuations by present-day and future experiments. The region around the curve corresponding to the first transition encloses the range of possible values for the signal in this line. The upper limit of this curve is calibrated on the Antennae. This local merging has a very bright CO~(1-0) and a star formation rate comparable to M82. This results in a higher value of the $\mathcal{R}$ ratio and in a higher amplitude of the signal. As a lower limit, we used the IRAS~14378 galaxy. Similar ranges of uncertainty can be drawn around the other lines, according to the limits presented in Fig.~\ref{fig:COplot}.

The contribution from dust emission in star-forming galaxies falls rapidly towards lower frequencies, where the CO signal seems to dominate. However, in this range, the foregrounds from compact radio sources \citep[including AGN, quasars, BL Lac objects, afterglows of gamma ray bursts, etc., see e.g.][]{dezotti05} is very relevant. Unfortunately, the clustering properties of such sources are very poorly known, though several authors showed that the Poisson power is likely to be largely dominant over the correlation one \citep{toffolatti98,gonzalez05}. The Poisson power, however, can be strongly reduced if the bright sources are excised from the map according to some flux cutoff. We show this in Fig.~\ref{fig:clradio}: here we consider the radio source counts at 33~GHz, from the model of \citet{dezotti05}. Integrating the counts we can estimate the Poisson contribution to the power spectrum. We compute it for several values of the flux cutoff and we find that the signal from the radio sources can be reduced to the level of the CO~(1-0) line emission if a $S_{\mathrm{cut}}\simeq0.1-1$~mJy is applied. The same cut does not affect the CO contribution significantly, since at this frequency both its Poisson and correlation term are generated by much weaker sources. There are two important caveats that have to be considered here. First, the main limitation to such cleaning procedure is the confusion noise. To reduce the radio signal to the CO level, we need to apply a flux cutoff on the order of $0.1-1$~mJy. According to the counts curve of \citet{dezotti05}, this corresponds to a density of $\sim20-200$ sources per square degree. Therefore the confusion noise should not be an issue, even at very low fluxes. Second, as mentioned above, the correlation term of the radio sources is considered negligible compared to the Poisson power. This is true, however, if the sources are not subtracted down to very low values of the flux \citep{toffolatti98}. In this case the contribution of correlation from weak, unresolved objects might become important, and it is necessary to have an accurate modeling of their clustering. 

The galactic foregrounds are also independent of the spectral resolution, so they can in principle be separated from the CO emission lines. 
Moreover, their angular power spectrum decreases rapidly towards the higher multipoles discussed here \citep{tegmark00}, especially at high galactic latitudes. The contribution from rotating dust is also characterized by a continuum spectrum \citep{dlaz98} and shows a peak around $20-30$~GHz in our galaxy \citep{deoliveira02,finkbeiner04}, which will be shifted towards lower frequencies in higher-redshift objects.

To conclude this section, we recompute the correlation term for the CO~(1-0) line using an alternative approach, based on a simple Press-Schechter distribution of the star-forming halos emitting in CO lines. This allows us to obtain a parallel constrain to compare with the estimates of our merging model, in particular at the lower frequencies corresponding to the higher redshifts.\\
Let us assume first that the star formation rate in halos scale with their mass
\begin{equation}
\dot{M}_{\star}=\mathcal{C}(z)\cdot M,
\end{equation}
and the factor $\mathcal{C}(z)$ is obtained by normalizing to the cosmic star formation rate density $\dot{\rho}_{\star}(z)$
\begin{equation}\label{hsnorm}
\dot{\rho}_{\star}(z)=\int dM\frac{dn}{dM}\mathcal{C}(z)M.
\end{equation}
The integration must be limited to the halos that are actually able to form stars. We use the fit of $\dot{\rho}_{\star}$ proposed by \citet{hernquistspringel03}, which uses a set of cosmological simulations up
to very high redshifts. These simulation were performed with a value of $\sigma_8^{\rm old}=0.9$, which is noticeably higher than the value adopted in this paper ($\sigma_8^{\rm new}=0.74$).  Therefore, before computing the normalization in Eq.~(\ref{hsnorm}) we rescale their fit taking this difference into account. Starting from Eq.~(19) in \citet{hernquistspringel03}, this rescaling can be written as
\begin{equation}
\frac{\dot\rho(\sigma_8^{\rm old},z)}{\dot\rho(\sigma_8^{\rm new},z)}
\approx\frac{\sigma_8^{\rm old}}{\sigma_8^{\rm new}} \exp\left\{-\frac{1}{2}\left(\frac{\delta_{\rm cr}}{D(z)\sigma_4^{\rm old}}\right)^2\left[1-\left(\frac{\sigma_8^{\rm old}}{\sigma_8^{\rm new}}\right)^2\right]\right\}
\end{equation} 
where $\sigma_4=\sigma(M_4)$ and $M_4$ is the mass corresponding to a virial temperature of $10^4$~K. This fit is accurate to better than $20-25\%$ for the redshifts corresponding to the peak of the signal and becomes even more precise at higher redshift. In either case, the error is much less than the other uncertainties of the model. \\
Fig.~\ref{fig:freqdep_nu-3} shows the result for the first CO transition obtained using this simple model: the estimate is a factor of a few below the one obtained with the merging approach. The dependence of this normalization on the actual value of $\sigma_8$ might open interesting possibilities of estimating the value of such a parameter from the amplitude of the signal.

\begin{figure}
  \resizebox{\hsize}{!}{\includegraphics{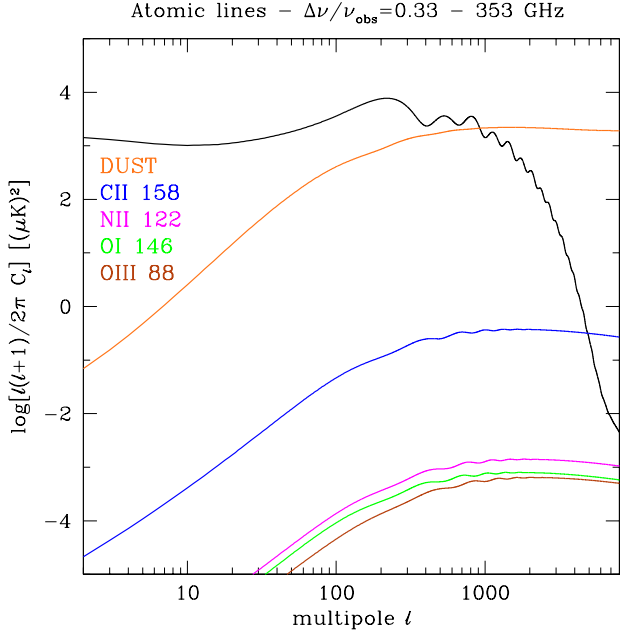}}
  \caption{The correlation signal of the most significant atomic emission lines, for a spectral resolution $\Delta\nu/\nu_{\mathrm{obs}}=0.33$ at 353~GHz, corresponding to one of the PLANCK HFI's detectors. The black line is the primordial signal of the CMB, the orange line the signal from dusty merging star-forming galaxies. Other colors identify the lines as indicated by the labels. See Table~\ref{tab:ATOM} for details.}
  \label{fig:cl353_atom}
\end{figure}

\begin{figure}
  \resizebox{\hsize}{!}{\includegraphics{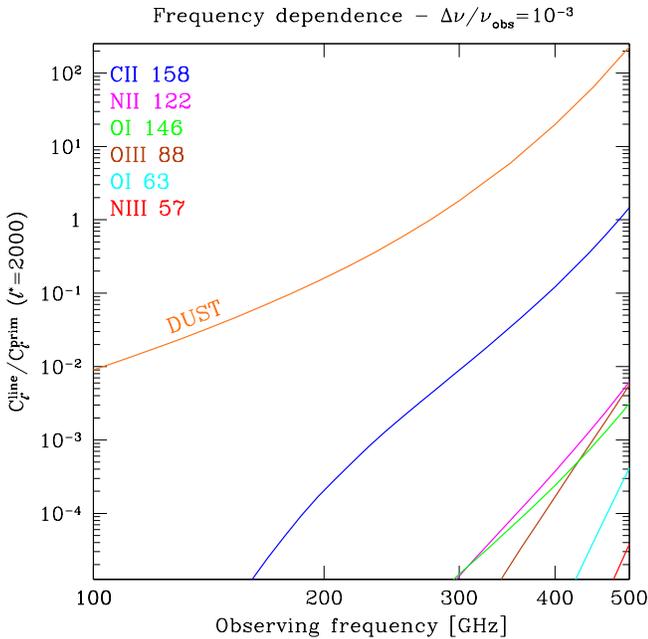}}
  \caption{The ratio of the correlation signal from atomic and ionic line emission to the primordial CMB signal at $l=2000$ as a function of frequency  and for a spectral resolution $\Delta\nu/\nu_{\rm obs}=10^{-3}$. The orange line is the signal from dust emission in merging galaxies. The CII line clearly dominates the other species, but is below the dust signal even at this spectral resolution.}
  \label{fig:freqdepATOM}
\end{figure}

\section{Angular fluctuations from fine-structure lines of atom and ions}\label{subsec:atom}
As mentioned in the introduction, the first generation of stars enriched the interstellar medium with significant amounts of metals, mainly carbon, oxygen, and nitrogen. The fine-structure transitions of these species are characterized by higher resonant frequencies with respect to the CO molecule. Therefore, their emission lines are more relevant for the high-frequency instruments, like PLANCK HFI, SPT, ACT, and ALMA. 

In an earlier paper, \citet{suginohara99} presented estimations of the collisional emission in
fine structure lines of these species. Assuming a fraction of $10^{-2}$ the solar abundance of those species, and assuming that roughly half of them are in environments with super-critical densities, they find that this emission should introduce an average distortion at the level of $10^{-6}$, whereas on scales of 1~$h^{-1}$Mpc, the relative intensity fluctuations should be a factor of 10 larger.

In this section we compute the signal due to the clustering of the sources responsible for the line emission. We repeat the approach used for the CO lines. To calibrate the $\mathcal{R}$ ratio for the fine-structure transitions of atoms and ions, we use the results of \citet{malhotra01}, who observed 60 star-forming galaxies with ISO-LWS. They especially selected the objects with a smaller angular size than the instrument beam, therefore the line fluxes presented in their paper refers to the total line emission from such galaxies. As for the low-redshift sample of CO lines, we obtain the FIR luminosities of these objects from the \emph{IRAS revised bright galaxies sample} \citep{sanders03}. 
For each of the 7 species presented in their work, we computed the geometric average of the $\mathcal{R}$ ratio in the sample as
\begin{equation}
\bar{\mathcal{R}}=\left[\prod_{i=1}^N \mathcal{R}_i\right]^{1/N}.
\end{equation}
The resulting numbers are listed in Table~\ref{tab:ATOM}.

\begin{table}
\caption{Average $\mathcal{R}$ for the 7 transitions of the sample from \citet{malhotra01}.}
\label{tab:ATOM}
\centering
\begin{tabular}{lcc}
\hline
\hline
Species	& Wavelength [$\mu$m]	& Average $\mathcal{R}$ \\
\hline
CII		& 158  			& $\bf6.0\ex{6}$	\\
OI		& 145  			& $\bf3.3\ex{5}$	\\
NII		& 122  			& $\bf7.9\ex{5}$	\\
OIII		& 88   			& $\bf2.3\ex{6}$	\\
OI		& 63   			& $\bf3.8\ex{6}$	\\
NIII		& 57   			& $\bf2.4\ex{6}$	\\
OIII		& 52   			& $\bf3.0\ex{6}$	\\
\hline
\end{tabular}
\end{table}

In Fig.~\ref{fig:cl353_atom} we show the correlation signal of these lines at 353~GHz and for a spectral resolution $\Delta\nu/\nu_{\mathrm{obs}}=0.33$, close to the performance of one of PLANCK HFI's channels. In the same figure we also show the correlation term from star-forming merging dusty galaxies computed in our previous paper \citep{righi08}; it is clear that this signal is still the dominant one.  Also in this case, however, it is possible to increase the amplitude of the correlation term by using very good spectral resolutions. We show this in Fig.~\ref{fig:freqdepATOM}, for a spectral resolution $\Delta\nu/\nu_{\rm obs}=10^{-3}$: the relative contribution of the CII~158 $\mu$m line increases with frequency when compared with the continuum emission from dust in merging star-forming galaxies. Other lines, plotted in the same figure, have a much lower amplitude and are therefore negligible.

\subsection{Comparison with similar works}
An estimate of the fluctuations arising from atomic line emission was presented by \citet{suginohara99}. When looking at the CII line at $z=10$, we find that our results predict emission amplitudes that are significantly (a factor of $\sim10^2$) less than theirs. Most of this disagreement is due to the higher value adopted for the density of ions emitting in the line (which depends on the assumed metallicity and ionization/excitation fractions), while having a different set of cosmological parameters introduces an offset of a factor of a few. In our work, the actual abundance of ions in the upper state is provided by the calibration of our model with observational data.

\begin{figure}
  \resizebox{\hsize}{!}{\includegraphics{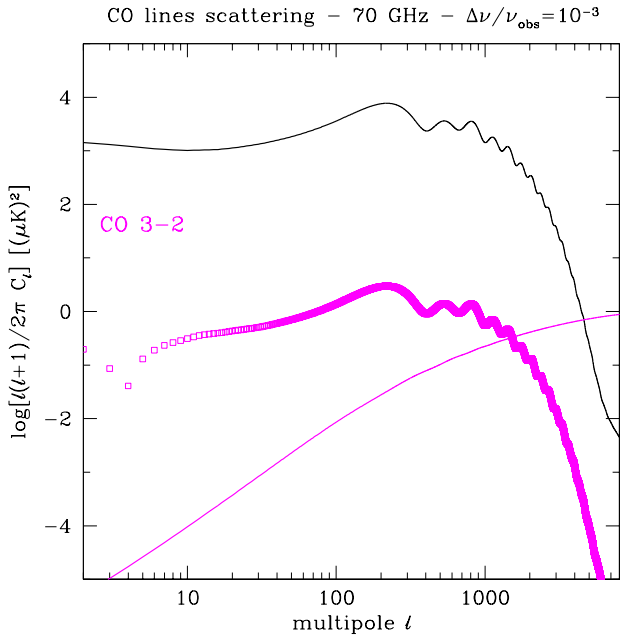}}
  \caption{The effect of the scattering of the CO~(3-2) line at 70~GHz (squares), according to the model of \citet{basu04}. The solid line is the correlation term from the emission in the same transition, for a spectral resolution $\Delta\nu/\nu_{\mathrm{obs}}=10^{-3}$.}
  \label{fig:cl_scat}
\end{figure}

\citet{basu04} analyzed the effect on the CMB primordial spectrum induced by the resonant scattering of heavy elements in the intergalactic gas, expelled by supernova explosions, galactic winds, and jets. They demonstrate that this effect is the sum of a damping of the original fluctuations plus the generation of new anisotropy. They obtained a very simple analytical expression for the change induced in the angular power spectrum on small scales
\begin{equation}
\delta C_l\simeq-2\tau_{\mathrm{scat}}C_l,
\end{equation}
where $\tau_{\mathrm{scat}}$ is the scattering optical depth of the line. The linear scaling of $\delta C_l$ with the optical depth produces quite strong effect even for a low value of $\tau$. In Fig.~\ref{fig:cl_scat} we plot the $\delta C_l$ for the third transition of CO and compare it with the correlation signal of the emission from the same line. For 1\% solar abundance, the scattering optical depth of this line at 70~GHz is $\tau_{3-2}=1.94\ex{-4}$.  It therefore introduces a change in the primordial power spectrum on the order of $\sim4\times10^{-4}$. This term, however, dominates on large scales, while it drops towards higher multipoles, following the decrease in the primordial spectrum, where the correlation signal dominates.

More recently, \citet{chm07,chm08} considered the pumping effect of the ultraviolet background on the OI 63 $\mu$m line and the resulting distortion on the CMB. They compute the correlation signal associated with the clustering of the first star-forming objects. The amplitude they find is on the level of $10^{-6}$ ($\mu$K)$^2$, and it is therefore lower than the signal described here. However, this effect would show up at higher frequencies ($\nu\sim400-700$~GHz), well within the spectral coverage of ALMA, whose sensitivity should be able to put constraints on this effect. We notice that the physical environments where both the collisional emission and the UV-induced emission take place are very similar, and therefore both effects constitute different probes for the same scenarios.

\section{Conclusions}\label{sec:concl}

The same star-forming activity that causes reionization of the IGM at $z\sim 10$ leaves an imprint on the CMB by means of several  mechanisms (e.g., Thomson scattering on the ionized gas, IR emission from dust particles that reprocess UV radiation, resonant scattering on metals produced by the first stars, etc.). In this work, we have focused on the impact that emission on molecular and atomic and ionic lines have on the angular power spectrum of the CMB. We put particular emphasis on the emission of CO rotational lines, since their effect is particularly strong in the $20-60$~GHz frequency range, although we have also
considered the collisional emission of lines corresponding to species like CII, OI, NII, and OIII.

If the star formation activity follows the halo merging history, then the spatial distribution of molecules and metals should closely resemble that of the halos where they were produced; therefore the emission pattern generated by these molecules and metals should show a similar clustering pattern. The advantage of looking at the emission of particular lines is that each observing frequency probes a given redshift shell, whose width is determined by the experimental spectral resolution ($\Delta z/z\sim\Delta\nu/\nu_{\mathrm{obs}}$). The anisotropy introduced by the clustering of the sources will be optimally measured if both the angular and the spectral resolutions of the observing experiment are able to spatially resolve those scales corresponding to the clustering ($\sim 15-25 $ $h^{-1}$ Mpc, in comoving units). On top of these clustering-induced fluctuations, we have to add the fluctuations associated to the Poisson nature of the sources, which contribute down to much smaller scales (the typical source size) and for which further improvements on $\Delta \nu / \nu$ result in a larger amount of measured anisotropy. In this context, all other signals that contribute in similar spectral ranges (dust IR emission, synchrotron radiation, intrinsic CMB, etc.) show a very different behavior with changing spectral resolution, because they merely remain constant. This will provide a powerful tool for distinguishing between the fluctuations characterized in this paper from all the rest {\em and} an efficient way to perform tomography of reionization at different frequency bands.

Due to the large number of uncertainties when building a theoretical model for the collisional emission on molecular, atomic, and ionic lines, we used existing measurements at low and high redshift to calibrate our model. Our approach is similar to \citet{righi08}, where our halo merging history model assigns a value of the the star formation rate to a given object and where this rate is assumed to follow a linear scaling with the luminosity under study (in this case, the luminosity in the various lines considered). For the CO rotational lines, we collected a sample of local and high-redshift objects of different sorts (including local mergings, bright infrared sources, AGN, QSOs, and radiogalaxies) to calibrate the $\mathcal{R}=L/\dot{M}$ ratio. We have considered, for each transition, a broad range of possible values to give an upper and lower limit estimate of the expected amplitude of the fluctuations.

Our results show that CO emission is already stronger than dust emission in the frequency range $10-70$~GHz if $\Delta \nu/\nu=10^{-3}$. Moreover, if one makes use of the different behavior of the CO signal
with the observing spectral resolution, then it should be possible to project out continuum signals like CMB, radio, and dust emission with greater confidence. In particular, the frequency interval ranging from 30 up to 50~GHz is of special interest for the first three CO rotational lines, since they probe the redshift range $z\simeq6-10$, crucial for our understanding of reionization. Atomic and ionic emission lines are more important at higher frequency, but only the CII 158~$\mu$m gives a non-negligible contribution to the amplitude.

This brings us to the conclusion that the emission on CO and CII lines studied in this paper provides a new window into reionization, constituting a bridge between the low-frequency observations pursuing the HI 21~cm
fluctuations in the radio range, and the high-frequency observations targeting fine structure lines in metals and ions like CII, OI, or OIII \citep{basu04,chm07,chm08}. These three windows should allow
tomography of the same cosmological epoch to be performed, but should be affected, in general, by very different contaminants and systematics. Therefore, the combination of these three perspectives should provide a consistent picture of an epoch that, to date, has remained hidden to observations.

\begin{acknowledgements}
We thank F.~Combes, L.~Tacconi and T.~Wilson for helpful discussions. MR is grateful to D. Docenko and U. Maio for their useful suggestions.
\end{acknowledgements}

\bibliographystyle{aa}
\bibliography{0199.bib}

\appendix
\section{The induced angular power spectrum}\label{sec:lineformal}
\subsection{The measured intensity}
In this section we compute the angular power spectrum induced on the CMB by the emission of these lines. We follow the line of sight approach (los) already outlined in Appendix A of \citet{righi08} and in \citet{chm08}, although here we place more emphasis on particular aspects related to the observing frequency resolution. Let $B(\vnh,\vnh')$ be the PSF of the observing beam on the sphere unit vector $\vnh'$ while pointing along $\vnh$. Let $\nu_{\mathrm{obs}}$ also be the observing frequency of the experiment that measures the emission of a line of rest frequency $\nu$ and therefore probes the universe at a redshift
$1+z = \nu/\nu_{\mathrm{obs}}$. The spectral intensity introduced by such line reads
\begin{eqnarray}\label{eq:Inu1st}
\Delta I_{\nu} &=& \int dr \int d\vnh\; B(\vnh,\vnh')\int d\nu'\; \phi_{\nu',\mathrm{instr}} \; \times\nonumber\\
&&\int d\vy\, dL_{\nu'(1+z)}\frac{dn(\vy)}{dL_{\nu'(1+z)}} W(\vy - \vr')
j_{\nu'(1+z)}\; a^3.
\end{eqnarray}
Here, $\phi_{\nu', \mathrm{instr}}$ is the frequency response of the experiment, and
$dn/dL_{\nu'(1+z)}$ the luminosity function of halos hosting the
metal/ionic species responsible for the line emission under study. The
emissivity on the line at the center of the halo is given by $j_{\nu'(1+z')}$,
which is diluted by cubic power of the scale factor $a$. The function
$W(\vy - \vr')$ provides the profile of the density distribution of the
species in the halo. The vector $\vr'$ is determined by $\vnh'$ and the
frequency $\nu'$ providing the effective distance to the observer: $\vr' =
r'(\nu ')\;\vnh'$. At this point, all spatial coordinates are physical. As
shown in \citet{righi08}, the convolution in Eq.~(\ref{eq:Inu1st}) can be
rewritten in terms of the source luminosity
\begin{eqnarray}\label{eq:Inu2nd}
\Delta I_{\nu}&=&\int dr \int d\vnh\; B(\vnh,\vnh')\int d\nu'\;
\phi_{\nu',\mathrm{instr}} \; \times\nonumber\\
&&\int dL_{\nu'(1+z)} \frac{d{\tilde n}(\vr')}{dL_{\nu'(1+z)}} \frac{a^3\;L_{\nu'(1+z)}}{4\pi}.
\end{eqnarray}
The new number density $d{\tilde n}/dL_{\nu'(1+z)}$ is just the previous one smoothed on scales corresponding to the typical halo size. The spectral luminosity can be expressed as a product of the bolometric luminosity on the line and the emission spectral profile, $L_{\nu'(1+z)}=L_{\mathrm{bol}}\psi_{\nu'(1+z)}$. We approximate this profile as a top hat function of the thermal width of the line,  ($\psi_{\nu'(1+z)} = 1/(\Delta \nu)_{\mathrm{th}}$ if $|\nu'(1+z) - \nu| < (\Delta \nu)_{\mathrm{th}}$, $\psi_{\nu'(1+z)} = 0$ otherwise), and assume that this width will be much smaller than the instrumental one ($(\Delta \nu)_{\mathrm{th}}/\nu \ll (\Delta \nu)_{\rm instr}/\nu_{\mathrm{obs}}$). This enables the replacement of the integral along $r$ by $cH^{-1}(z) (\Delta \nu)_{\mathrm{th}}/\nu$, yielding
\begin{eqnarray}\label{eq:Inu3rd}
\Delta I_{\nu}&=&\int d\vnh\; B(\vnh,\vnh')\int d\nu'\; \phi_{\nu',\mathrm{instr}}
\; \times\nonumber\\
&&\int dL_{\nu'(1+z)} \frac{dn(\vr')}{dL_{\nu'(1+z)}} \frac{a^3\;{\tilde
L}_{\nu'(1+z)}}{4\pi}\; cH^{-1}(z),
\end{eqnarray}
where $\tilde{\lnu}\equiv L_{\mathrm{bol}} / \nu$. This cancels any dependence on $(\Delta \nu)_{\mathrm{th}}$. Note that the halo number density is evaluated at $\vr'$ and that the integral along the line of sight is carried out under the frequency response of the instrument:
\begin{eqnarray}\label{eq:Inu4th}
\Delta I_{\nu}&=&\int dr'\; {\cal P}(r-r') \;\int d\vnh\; B(\vnh,\vnh')\;
\times\nonumber\\
&&\int dL_{\nu'(1+z)} \frac{dn(\vr')}{dL_{\nu'(1+z)}} \frac{a^3\;{\tilde L}_{\nu'(1+z)}}{4\pi}\; cH^{-1}(z).
\end{eqnarray}
The product ${\cal P}(r-r')\equiv \partial \nu' / (\partial r') \;\phi_{\nu', \mathrm{instr}} $ constitutes the radial profile of the los integration. We next attempt to show the impact that the shape of the radial (and angular) responses of the experiment have on the measured intensity. If we convert $\vnh'$ into the transversal spatial component by introducing $r'^2$, Eq.~(\ref{eq:Inu4th}) reads like a convolution
\begin{equation}\label{eq:Inu5th}
\Delta I_{\nu}(\vr) = \int d\vr'\;{\cal B}(\vr-\vr')   \;{\tilde
I}_{\nu'(1+z)} (\vr'),
\end{equation}
with ${\cal B}(\vr-\vr') \equiv {\cal P}(r-r')/r'^2\;$ and
\begin{equation}
{\tilde I}_{\nu'(1+z)}(\vr') \equiv cH^{-1}(z)\;\int
dL_{\nu'(1+z)}\frac{a^3\;\tilde{\lnu}_{'(1+z)}}{4\pi}\;\frac{d{\tilde
n(\vr')}}{dL_{\nu'(1+z)}}.
\label{eq:inudef}
\end{equation}
Here, ${\cal B}(\vr-\vr')$ represents the 3D window function of the instrumental response and is responsible for some smoothing of the original signal ${\tilde I}_{\nu'(1+z)} (\vr')$. This becomes more obvious if we rewrite the convolution in Fourier space
\begin{equation}
\Delta I_{\nu} (\vr)= \int \dkthree\; \exp{(-i\vk\cdot \vr)}\; {\cal
B}_{\vk}  \; {\tilde I}_{\nu'(1+z)} (\vk ).
\label{eq:Inu6th}
\end{equation}
If the instrument is not sensitive to small scales (due to a poor angular {\em or} spectral resolution), then at large $\vk$ the ${\cal B}_{\vk}$ will vanish, suppressing the contribution of ${\tilde I}_{\nu'(1+z)} (\vk)$ at the same $\vk$'s. This is also reflected in the difference scaling of the power spectra versus the spectral resolution of the experiment. Indeed, if we compute the power spectrum of $\Delta I_{\nu} (\vr)$ in terms of the halo power spectrum, we obtain
\begin{equation}
|\Delta I_{\nu, \vk}|^2 \propto |{\cal B}_{\vk}|^2 \; 
P_h(k)\;\biggl(cH^{-1}(z) \frac{\tilde {L}_{\nu'(1+z)}} {4\pi}
\frac{d\bar{n}}{dL_{\nu'(1+z)}} \biggr)^2,
\end{equation}
with $d\bar{n}/d\lnu$ the {\em average} number density of halos per unit luminosity. If the halos are Poisson-distributed, then $P_h(k) \propto 1/ (d\bar{n}/dL_{\nu'(1+z)})$ up to very large $k$-s, and the high-$k$ integral is actually limited by the window function ${\cal B}_{\vk}$ (as long as $k$ corresponds to scales larger than the source size). In contrast, if halos are clustered so that their power spectrum is proportional to the linear matter density power spectrum $P_m(k)$, then we find that $P_m(k) \rightarrow 0$ for some large $k$, and the increase in spectral/angular resolution of the experiment will not make any difference at this high $k$ range. {\em Therefore, improving the spectral resolution helps for increasing the Poisson
contribution, but does not change the contribution from the clustering term beyond some large $k_c = 2\pi/L_c$, with $L_c$ the typical clustering
scale}.

\subsection{The angular power spectrum}
Our starting point is Eq.~(\ref{eq:Inu4th}) above, which can easily be
converted in comoving units by absorbing the $a^3$ factor within the
definition of comoving number density. An $a$ factor must then be added for the line
element $dr$, but this cancels out with the $a^{-1}$ for the profile $\mathcal{P}$. To compute the angular power spectrum, we have to
characterize the clustering properties of the emitting halos. We recall that
\begin{equation}
\frac{dn}{dL_{\nu'(1+z)}}=\int dM\,\frac{dn}{dM}\,G(M,L_{\nu'(1+z)}),
\end{equation}
where the function $G(M,L_{\nu'(1+z)})$ provides the fraction of halos present in the mass interval $[M,M+dM]$ (given by the mass function $dn/dM$) that have recently experienced a major merger (see \citet{righi08}), and hence given rise to significant emission in the line(s) of interest. The spatial correlation function of halos in the mass range $[M,M+dM]$, $n_{\mathrm{h}}(M,\vx)$ reads
\[
\langle n_{\mathrm{h}}(M_1,\vec{x}_1)\,n_{\mathrm{h}}(M_2,\vec{x}_2)\rangle=\frac{dn}{dM}(M_1,\vec{x}_1)\,\frac{dn}{dM}(M_2,\vec{x}_2)
\]
\[
\phantom{xxxxxxx}+\,\frac{dn}{dM}(M_1,\vec{x}_1)\,\delta_D^3(\vec{x}_1-\vec{x}_2)\,\delta_D(M_1-M_2)
\]
\[
\phantom{xxxxxxx}+\,\frac{dn}{dM}(M_1,\vec{x}_1)\frac{dn}{dM}(M_2,\vec{x}_2)
\]
\begin{equation}
\phantom{xxxxxxx}\times\, b(M_1,z[x_1])\,b(M_2,z[x_2])\,\xi_{\mathrm{m}}(\vec{x}_1-\vec{x}_2).
\end{equation}
The symbol $\delta_D$ stands for Dirac delta and $b(M,z)$ is the mass and redshift dependent bias factor that relates the halo and the matter linear
correlation function $\xi_{\mathrm{m}}$ \citep{mowhite02}. The first term on the righthand side accounts for the Poissonian fluctuations in the number
counts, whereas the second term accounts for the dependence of the halo number density on the environment.

Having this present, just as in Appendix A of \citet{righi08}, it is possible to write the angular correlation function of the intensity fluctuations as
\begin{equation}
\langle\Delta I_{\nu}(\vnh_1)\,\Delta I_{\nu}(\vnh_2
)\rangle=\sum_l\frac{2l+1}{4\pi}\left(C_l^P+C_l^C\right)P_l(\vnh_1
\cdot \vnh_2),
\end{equation}
with $P_l(\vnh_1 \cdot \vnh_2)$ the Legendre polynomia of order $l$ and where $C_l^P$ and $C_l^C$ are the Poissonian and the correlation terms of the $l$-th angular power spectrum multipole. Then, $C_l^P$ can easily be found to be
\begin{eqnarray}\label{eq:Cl_p}
C_l^P&=& \left[ r^2 \;\frac{(\Delta
\nu)_{\mathrm{instr}}}{\nu_{\mathrm{obs}}}\right]^{-1} \times\nonumber\\
&&\int dL_{\nu'(1+z)}\;cH^{-1}(z) \left(\frac{\tilde{L}_{\nu'(1+z)}} {4\pi}\right)^2 \frac{d\bar{n}}{dL_{\nu'(1+z)}}.
\end{eqnarray}
On the other hand, the correlation term can be expressed in terms of a $k$-space integral of the initial scalar metric power spectrum
$P_{\psi}(k)$ times a squared transfer function, just as in \citet{cmbfast},
\begin{equation}
\label{correl}
C_l^C=\frac{2}{\pi}\int k^2dk\,P_{\psi}(k)\,|\Delta_l (k)|^2,
\end{equation}
with the transfer function $\Delta_l (k)$ given by
\begin{equation}
\Delta_l(k)=\int dr\,j_l(kr)\; {\cal P}(r)\left[S(r)\,\delta_k\right].
\end{equation}
In this equation, $j_l(x)$ is the spherical Bessel function of order $l$, ${\cal P}(r)$ is the instrumental profile function as defined after
Eq.~(\ref{eq:Inu4th}), $\delta_k$ is the $k$-mode of the dark matter density contrast, and the function $S(r)$  is defined by
\begin{eqnarray}
S(r)&\equiv&\int dL_{\nu'(1+z)}\,dM\,G(M,L_{\nu'(1+z)})\nonumber\\
&&\times\,\frac{dn}{dM}\,\frac{{\tilde
L}_{\nu'(1+z)}}{4\pi}\,cH^{-1}(z)\,b(M,z[r]).
\end{eqnarray}

\section{Observational method}\label{sec:method}
In this Appendix we propose a simple method for extracting the line-induced
signal from a continuum at a given frequency, by the use of observations
at different spectral resolutions. 

Our model for the measured signal (in either real or Fourier space) is given
by an array of $N$ channels at different spectral resolutions ($\delta_i
\equiv (\Delta \nu)_i / \nu$, $i=1,N$):
\begin{equation}
\tilde{s}_i = \alpha f(\delta_i) + C + N_i.
\label{eq:models1}
\end{equation}
The vector $\delta_i$ should sweep the relevant spectral resolution range
shown in Fig.~\ref{fig:specdep}, ($\delta_i \in [0.1,
5\times10^{-4}]$).
The value of $\alpha$ provides the amplitude of the line-induced emission,
whereas the function $f(\delta_i)$ contains its dependence with respect to
the
spectral resolution. {\it A priori}, this function can be characterized if
the
redshift of the line-induced emission is known. Signal $C$ refers to all
components that do not show an explicit dependence of spectral resolution
(e.g., intrinsic CMB, dust, synchrotron, etc.) and whose value should remain
constant (at least down to a few percent) in every channel. Finally, $N_i$
refers to the instrumental noise in
each channel, and its covariance matrix is assumed to be characterized.
With this in hand, it is convenient to take differences between measurements
in different channels having distinct values for $f(\nu)$,
\begin{equation}
d_{ij}\equiv \tilde{s}_i - \tilde{s}_j = \alpha \left(f(\delta_i) -
f(\delta_j)\right) + N_{ij},
\label{eq:diff1}
\end{equation}
where $N_{ij}$ denotes a combination of the noise from both channels.

An optimal estimate for the line-emission amplitude can be obtained by
minimizing the quantity
\begin{eqnarray}
\chi^2 &=& \sum_{ij,lm} \left[ d_{ij} -  {\tilde \alpha} \left(f(\delta_i) -
                                        f(\delta_j)\right) \right] {\cal N}_{ij,lm}^{-1}\times\nonumber\\
             &&\left[ d_{lm} -  {\tilde \alpha} \left(f(\delta_l) -
                                        f(\delta_m)\right) \right],
\label{eq:chis1}
\end{eqnarray}
i.e., by finding the value of ${\tilde \alpha}$ that satisfies $\partial
\chi^2 /\partial {\tilde \alpha} = 0$. The covariance matrix of the combined
channel noise is given by $ {\cal N}_{ij,lm}$. The minimization yields
\begin{equation}
{\tilde \alpha} = \frac{ \sum_{ij,lm} d_{ij} {\cal N}_{ij,lm}^{-1}
\left(f(\delta_l) - f(\delta_m)\right)}{\sum_{ij,lm} \left(f(\delta_i) -
                                        f(\delta_j)\right) {\cal N}_{ij,lm}^{-1}
                        \left(f(\delta_l) -     f(\delta_m)\right) },
\label{eq:alphatilde}
\end{equation}
with a formal error
\begin{equation}
\sigma^2_{\tilde \alpha} = \frac{1}{\sum_{ij,lm} \left(f(\delta_i) -
                                        f(\delta_j)\right) {\cal N}_{ij,lm}^{-1}
                        \left(f(\delta_l) -     f(\delta_m)\right) }.
\label{eq:sgalphatilde}
\end{equation}

\end{document}